\documentclass[secnumarabic,superscriptaddress,amssymb, nobibnotes, aps, prl]{revtex4-2}

\usepackage{amsmath}
\usepackage{mathrsfs}  % use to write fancy letters (e.g. for free energy)
\usepackage{mathtools} % use to write := im math
\usepackage{bbold}    % use to write fancy 1 for unity matrix
\usepackage{braket}    % lets you use the bra, ket notation
\usepackage{gensymb}   % for \degree

\usepackage{graphicx}      	% import figures
\graphicspath{{figures/}}
\usepackage{multirow}       % merge rows and columns in table
\usepackage{here} 			% display figures exactly where they are in code
\usepackage{url}		% display urls in a nice way
\usepackage[plainpages=false,pdfpagelabels=true, linkcolor=cyan,breaklinks,colorlinks=true]{hyperref}  % hyperrefs
\usepackage{soul}                   % allows to cross out text
\usepackage[dvipsnames]{xcolor}		% text in different colors
\usepackage{siunitx}
\usepackage{xspace}
\usepackage{booktabs}    % for prettier lines in table

\newcommand{\ute}{UTe\textsubscript{2}\xspace}

\newcommand{\sro}{Sr\textsubscript{2}RuO\textsubscript{4}\xspace}
\newcommand{\Hct}{\ensuremath{H_{\rm c2}}\xspace}
\newcommand{\Tc}{\ensuremath{T_{\rm c}}\xspace}
\newcommand{\Tcs}{\ensuremath{T_{\rm c}}'s\xspace}

\begin{document}

\title{Single-Component Superconductivity in \ute at Ambient Pressure}%

\author{Florian Theuss}%
\affiliation{Laboratory of Atomic and Solid State Physics, Cornell University, Ithaca, NY 14853, USA}
\author{Avi Shragai}
\affiliation{Laboratory of Atomic and Solid State Physics, Cornell University, Ithaca, NY 14853, USA}
\author{Gael Grissonnanche}
\affiliation{Laboratory of Atomic and Solid State Physics, Cornell University, Ithaca, NY 14853, USA}
\affiliation{Kavli Institute at Cornell for Nanoscale Science, Ithaca, New York, USA}
\author{Ian M Hayes}
\affiliation{Maryland Quantum Materials Center, Department of Physics, University of Maryland, College Park, Maryland 20742, USA}
\author{Shanta R Saha}
\affiliation{Maryland Quantum Materials Center, Department of Physics, University of Maryland, College Park, Maryland 20742, USA}
\author{Yun Suk Eo}
\affiliation{Maryland Quantum Materials Center, Department of Physics, University of Maryland, College Park, Maryland 20742, USA}
\author{Alonso Suarez}
\affiliation{Maryland Quantum Materials Center, Department of Physics, University of Maryland, College Park, Maryland 20742, USA}
\author{Tatsuya Shishidou}
\affiliation{Department of Physics, University of Wisconsin-Milwaukee, Milwaukee, Wisconsin 53201, USA}
\author{Nicholas P Butch}
\affiliation{Maryland Quantum Materials Center, Department of Physics, University of Maryland, College Park, Maryland 20742, USA}
\affiliation{NIST Center for Neutron Research, National Institute of Standards and Technology, 100 Bureau Drive, Gaithersburg, Maryland 20899, USA}
\author{Johnpierre Paglione}
\affiliation{Maryland Quantum Materials Center, Department of Physics, University of Maryland, College Park, Maryland 20742, USA}
\affiliation{Canadian Institute for Advanced Research, Toronto, Ontario, Canada}
\author{B. J. Ramshaw}
\email{bradramshaw@cornell.edu}
\affiliation{Laboratory of Atomic and Solid State Physics, Cornell University, Ithaca, NY 14853, USA}
\affiliation{Canadian Institute for Advanced Research, Toronto, Ontario, Canada}

\date{\today}%

\begin{abstract}
\bf
The microscopic mechanism of Cooper pairing in a superconductor leaves its fingerprint on the symmetry of the order parameter. \ute has been inferred to have a multi-component order parameter that entails exotic effects like time reversal symmetry breaking. However, recent experimental observations in newer-generation samples have raised questions about this interpretation, pointing to the need for a direct experimental probe of the order parameter symmetry. Here, we use pulse-echo ultrasound to measure the elastic moduli of \ute in samples that exhibit both one and two superconducting transitions. We demonstrate the absence of thermodynamic discontinuities in the shear elastic moduli of both single- and double-transition samples, providing direct evidence that \ute has a single-component superconducting order parameter. We further show that superconductivity is highly sensitive to compression strain along the $a$ and $c$ axes, but insensitive to strain along the $b$ axis. This leads us to suggest a single-component, odd-parity order parameter---specifically the B$_{2u}$ order parameter---as most compatible with our data.
\end{abstract}

\maketitle
% \tableofcontents

%%%%%%%%%%%%% input introduction
\section{Introduction}

Definitive determinations of the superconducting pairing symmetry have been accomplished for only a handful of materials, among them the $s$-wave BCS superconductors and the $d$-wave cuprates \cite{tsuei_pairing_2000}. In some superconductors, such as \sro, debate over the pairing symmetry has persisted for decades despite ultra-pure samples and an arsenal of experimental techniques \cite{ghosh_thermodynamic_2021,rice_sr_1995,mackenzie_even_2017}. This is more than an issue of taxonomy: the pairing symmetry places strong constraints on the microscopic mechanism of Cooper pairing, and some pairing symmetries can lead to topological superconducting states \cite{sato_topological_2017}. 

The question of pairing symmetry is nowhere more relevant than in \ute, where in addition to power laws in thermodynamic quantities \cite{metz2019PointnodeGapStructure,ranNearlyFerromagneticSpintriplet2019,ishiharaChiralSuperconductivityUTe22023,kittaka2020OrientationPointNodes}, the most striking evidence for unconventional superconductivity is an extremely high upper critical field \Hct compared to the relatively low critical temperature \cite{ranNearlyFerromagneticSpintriplet2019,aokiUnconventionalSuperconductivityHeavy2019}. Remarkably, for some field orientations, the superconductivity re-emerges from a resistive state above $\sim$40 tesla and persists up to at least 60 tesla \cite{ran_extreme_2019}. This high \Hct constrains the spin component of the Cooper pair to be spin-triplet, which in turn constrains the orbital component of the Cooper pair to be odd under inversion (i.e. odd parity, such as a $p$ or $f$-wave state). However, there are many possible odd-parity order parameters and which one---or which pair, if \ute is a two-component superconductor as suggested \cite{hayes2021multicomponent}---manifests in \ute is unknown.

The primary question we address here is regarding the degeneracy of the orbital part of the superconducting order parameter. In addition to even ($s$ or $d$-wave) and odd ($p$ and $f$-wave) designations, order parameters can have multiple components: both conventional $s$-wave and high-\Tc $d_{x^2-y^2}$-wave order parameters are described by a single complex number, whereas the topological $p_x+i p_y$ state has two components, namely $p_x$ and $p_y$. Evidence for a two component order parameter in \ute stems from the presence of two distinct superconducting transitions in some samples, as well as the onset of time-reversal symmetry breaking at \Tc \cite{hayes2021multicomponent,weiInterplayMagnetismSuperconductivity2022}. Combined with the evidence for spin-triplet pairing, these observations have led to several proposed exotic, multi-component order parameters for \ute (see \autoref{tab:op}). These multi-component states can have a topological structure that could explain other experimental observations, such as the chiral surface states seen in STM \cite{Jiao2020}, or the anomalous normal component of the conductivity observed in microwave impedance measurements \cite{baeAnomalousNormalFluid2021}.

Claims of a multi-component order parameter are not without controversy, however. As the purity of the samples has increased, \Tc has shifted to higher values and the second transition has disappeared at ambient pressure \cite{rosaSingleThermodynamicTransition2022}. Previous work has suggested that two transitions arise due to inhomogeneity \cite{thomasSpatiallyInhomogeneousSuperconductivity2021a}, but the application of hydrostatic pressure splits single-\Tc samples into two-\Tc samples \cite{aokiMultipleSuperconductingPhases2020,braithwaiteMultipleSuperconductingPhases2019}, suggesting that two superconducting order parameters are, at the very least, nearly degenerate with one another.

\begin{figure}%
\includegraphics[width=0.75\columnwidth]{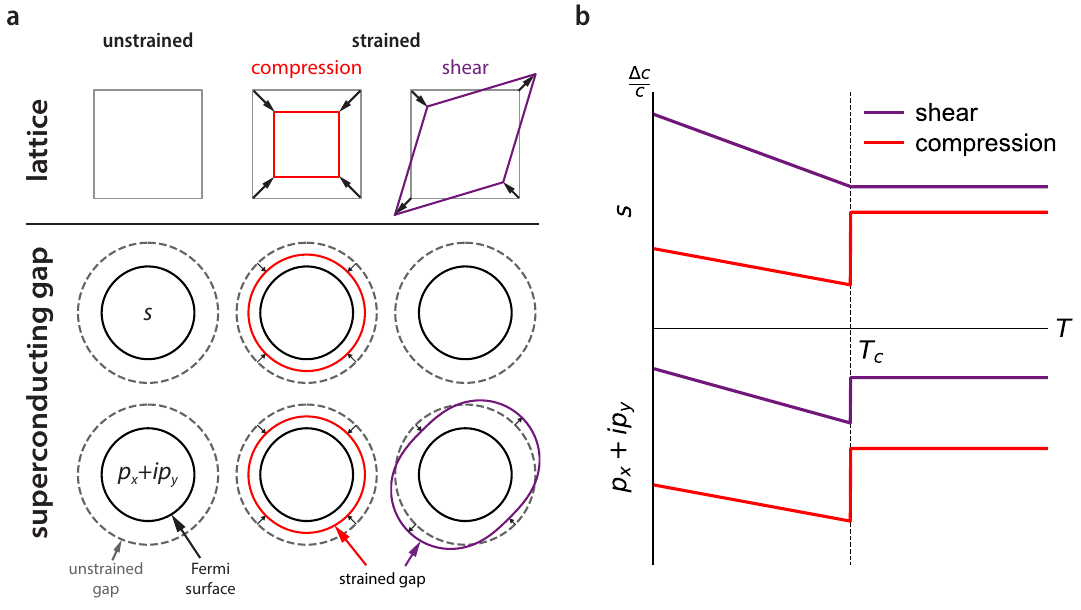}%
\caption{\textbf{The influence of strain on one- and two-component superconductors.} Panel (a) illustrates how two representative order parameters---single-component $s$-wave and two-component $p_x + i p_y$---respond to both compression and shear strain. Both gaps respond under compression (whether increasing or decreasing in magnitude depends on microscopic details). Only the two-component gap, however, couples to shear strain---here we illustrate the ``phase'' mode (see \citet{ghosh_thermodynamic_2021} for more details). Panel (b) shows the expected changes in elastic moduli across \Tc for one and two component order parameters. All superconductors have a discontinuity in their compressional moduli across \Tc, but only two-component superconductors have discontinuities in their shear moduli.}%
\label{fig:schematic}%
\end{figure}

The natural way to distinguish between single-component and two-component order parameters is to apply strain. Single-component superconductors have a single degree of freedom that couples to compression strains---the superfluid density---producing a discontinuity in the compressional elastic moduli at \Tc (see \autoref{fig:schematic}). They, however, have no such discontinuity in their shear moduli because shear strains preserve volume and thus do not couple to superfluid density. Multi-component superconductors, on the other hand, have additional degrees of freedom: the relative orientation of the two order parameters, as well as their relative phase difference. These additional degrees of freedom couple to shear strains, producing discontinuities in the shear moduli at \Tc. By identifying which elastic moduli have discontinuities at \Tc, one can determine whether a superconductor is multi-component without any microscopic knowledge of the Fermi surface or the pairing mechanism.

\begin{table}[]
	% \centering
	\begin{tabular}{c c c l}
		Dimensionality                    & Representation    &Shear discontinuity?         & Reference (E: experiment; T: theory)\\
		\toprule
		\multirow{8}{*}{One-Component}   & \multirow{2}{*}{$A_{u}$} & \multirow{2}{*}{No} & E: NMR \cite{matsumuraLargeReductionAaxis2023}\\
										  &                          &  & E: scanning SQUID \cite{iguchiMicroscopicImagingHomogeneous2023}\\
										  \cmidrule{2-4}
																			&	\multirow{1}{*}{\color{Red} $B_{2u}$}								&	\multirow{1}{*}{\color{Red} No} & E: {\color{Red} Ultrasound (this work)}\\
											\cmidrule{2-4}
		                                  &\multirow{4}{*}{$B_{3u}$} & \multirow{3}{*}{No} & T: DFT \cite{xuQuasiTwoDimensionalFermiSurfaces2019}\\
		                                %   &                          &  & T: Hund's-Kondo model\cite{hazra2023TripletPairingMechanisms}\\
		                                  &                          &  & E: NMR\cite{fujibayashiSuperconductingOrderParameter2022a,nakamineAnisotropicResponseSpin2021a}\\
		                                  &                         &   & E: scanning SQUID \cite{iguchiMicroscopicImagingHomogeneous2023}\\
										  \cmidrule{2-4}
										  &                         & \multirow{2}{*}{No} & E: specific heat \cite{rosaSingleThermodynamicTransition2022,thomasSpatiallyInhomogeneousSuperconductivity2021a}\\
										  &                         &  & E: uniaxial stress \cite{girod2022ThermodynamicElectricalTransport}\\
		\midrule
		\multirow{13}{*}{Two-Component}   & \multirow{2}{*}{$\{B_{1u}, A_{u}\}$} & \multirow{2}{*}{$c_{66}$}& E: microwave surface impedance \cite{baeAnomalousNormalFluid2021}\\
										  &                                     &  & E: specific heat, Kerr effect \cite{hayes2021multicomponent}\\
										  \cmidrule{2-4}
										  & \multirow{2}{*}{$\{B_{3u}, A_{u}\}$} & \multirow{2}{*}{$c_{44}$}& E: penetration depth \cite{ishiharaChiralSuperconductivityUTe22023}\\
										  &                                      & & E: NMR \cite{nakamine2021InhomogeneousSuperconductingState}\\       
										  \cmidrule{2-4}
										  & \multirow{2}{*}{$\{B_{1u}, B_{2u}\}$} & \multirow{2}{*}{$c_{44}$} & T: phenomenology analoguous to $^3$He \cite{machida2020TheorySpinpolarizedSuperconductors,machida2021NonunitaryTripletSuperconductivity}\\
										  & 					                & & E: specific heat \cite{kittaka2020OrientationPointNodes}\\
										  \cmidrule{2-4}
										  & \multirow{2}{*}{$\{B_{1u}, B_{3u}\}$}&\multirow{2}{*}{$c_{55}$} & T: phenomenology + DFT \cite{nevidomskyy2020StabilityNonunitaryTriplet}\\
									      & 					                &  & T: DFT \cite{ishizuka2019InsulatorMetalTransitionTopological}\\				
										  \cmidrule{2-4}
										  & \multirow{3}{*}{$\{B_{2u}, B_{3u}\}$}& \multirow{3}{*}{$c_{66}$}& T: DFT \cite{shishidou2021TopologicalBandSuperconductivity,choiCorrelatedNormalState2023}\\	
										  &  					                & & E: specific heat, Kerr effect \cite{hayes2021multicomponent,weiInterplayMagnetismSuperconductivity2022}\\
										  &  					               & & T: emergent $D_{4h}$ symmetry under RG flow \cite{shafferChiralSuperconductivityMathrmUTe2022}\\
										  \cmidrule{2-4}
										  & 	                               &  \multirow{3}{*}{Yes} & E: STM \cite{Jiao2020}\\
										  &                                    &                       & T: pair-Kondo effect \cite{hazraPairKondoEffectMechanism2022}\\
										  &  					               &                       & T: MFT of Kondo lattice \cite{changTopologicalKondoSuperconductors2023}\\
		% Two   & $\{A_{g}, A_{u}\}$ or $\{A_{g}, B_{3u}\}$  & \citet{kanasugiAnapoleSuperconductivity2022a} & theory; mixed-parity pairing \\%- could explain STM and Kerr and different magnetic interactions and different pressure dependence of SC OP's\\
		% Two   & $\{A_{g}, A_{u}\}$ or $\{A_{g}, B_{3u}\}$  & \citet{watanabeNonreciprocalMeissnerResponse2022} & theory; mixed-parity \\
		% Two   &mixed parity?         & \citet{ishizuka2021PeriodicAndersonModel} & theory - periodic Anderson model\\
		\bottomrule
	\end{tabular}
	\caption{\textbf{Proposed order parameters for \ute.} Proposed odd-parity order parameters for \ute, sorted by the number of components (dimensionality), their irreducible representation, and whether the proposed order parameter is based on an experimental observation or a theoretical proposal. Scenarios listed without a specific representation are compatible with any type of one or two-component order parameter. Based on symmetry alone, our work strongly constrains the order parameter to be of the one-component type. Using more quantitative arguments, we suggest a $B_{2u}$ order parameter.    }
	\label{tab:op}
\end{table}

\section{Results}

We use a traditional phase-comparison pulse-echo ultrasound technique to measure the temperature dependence of six elastic moduli in three different samples of \ute over a temperature range from about 1.3~K to 1.9~K. In particular, we measure all three compressional (i.e. $c_{11}$, $c_{22}$, and $c_{33}$) and shear (i.e. $c_{44}$, $c_{55}$, and $c_{66}$) moduli in one sample with two superconducting transitions (S3: $T_{c,1}\approx1.64$~K, $T_{c,2}\approx1.60$~K) and in two samples with a single \Tc (S1: $T_c\approx1.63$~K and S2: $T_c\approx1.70$~K). Ultrasound data in the normal state of \ute have been reported in \citet{ushida2023LatticeInstabilityUTe2}. Here, we focus on the superconducting transition. Details of the sample growth and preparation, as well as the experiment, are given in the Methods. 

\begin{figure}%
\includegraphics[width=0.95\columnwidth]{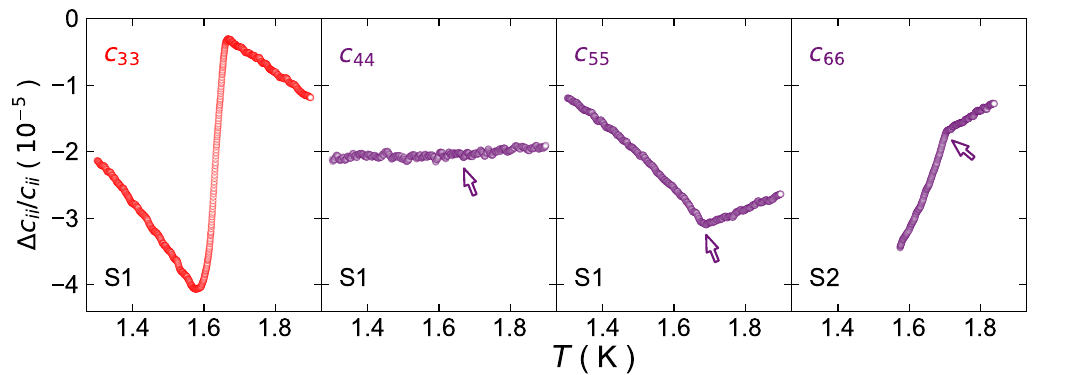}%
\caption{\textbf{Relative change in elastic moduli through \Tc for single-\Tc \ute.} A compressional elastic modulus, $c_{33}$, shows a sharp discontinuity of approximately 40 parts per million at \Tc, as expected for all superconductors. In contrast, the shear elastic moduli---$c_{44}$, $c_{55}$, and $c_{66}$---show only changes in slope at \Tc, consistent with a single-component superconducting order parameter. Empty arrows mark the superconducting transition for all shear moduli. $\Delta c_{ii}/c_{ii}$ is defined as $\left(c_{ii}(T)-c_{ii}(T_0)\right)/c_{ii}(T_0)$ where $T_0$ is the highest temperature in the plot.}%
\label{fig:singleTc}%
\end{figure}

\autoref{fig:singleTc} shows the relative changes in four elastic moduli across \Tc for the single-transition samples S1 and S2. We observe a single, sharp ($\approx$ 85 mK wide) discontinuity in the $c_{33}$ compressional modulus, as expected for all superconducting transitions. We observe no discontinuities in any of the shear elastic moduli to within our experimental resolution (a few parts in $10^7$, see S.I. for details).

\begin{figure}%
\includegraphics[width=0.95\columnwidth]{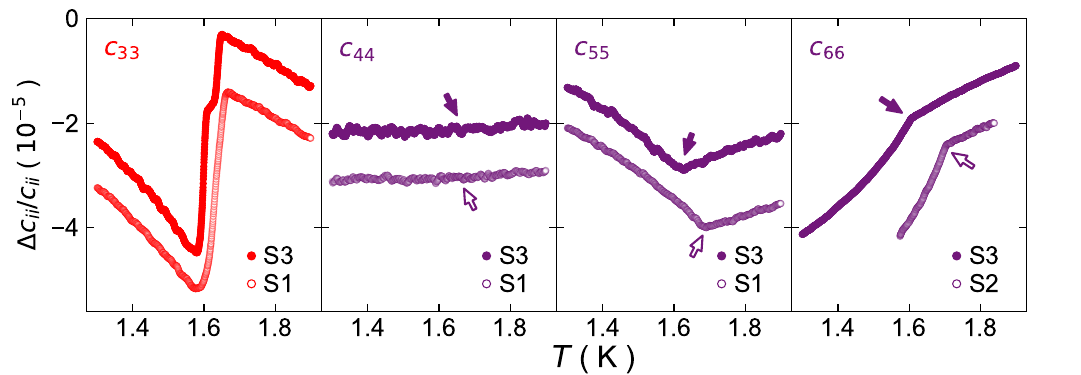}%
\caption{\textbf{Relative change in elastic moduli through \Tc for double-\Tc \ute.} The compressional elastic modulus $c_{33}$ shows two distinct discontinuities at \Tc, consistent with the two peaks we find in the specific heat of the same sample. The shear moduli, on the other hand, show no discontinuities and behave nearly identically to the shear moduli of the single-\Tc sample. Single (double) transition samples are shown with empty (filled) symbols. Empty (filled) arrows mark the superconducting transition for all shear moduli for single (double) transition samples. $\Delta c_{ii}/c_{ii}$ is defined as in \autoref{fig:singleTc}, and curves have been offset for clarity.}%
\label{fig:doubleTc}%
\end{figure}

\autoref{fig:doubleTc} shows the relative changes in the elastic moduli for sample S3 with a double superconducting transition (the single-\Tc data is reproduced here for comparison). We observe two distinct discontinuities in $c_{33}$ separated by approximately 40 mK. Subsequent specific heat measurements on the same sample show a similar ``double peak'' feature identified in other double-\Tc samples \cite{hayes2021multicomponent} (specific heat data is shown in the S.I.). Notably, we find the sum of the discontinuities in the double-\Tc sample to be of a similar size as the discontinuity in the single-\Tc sample. Additionally, the behaviour of the shear elastic moduli is nearly identical to that of the single-\Tc samples, again with no discontinuities at \Tc.

\begin{figure}%
\includegraphics[width=0.95\columnwidth]{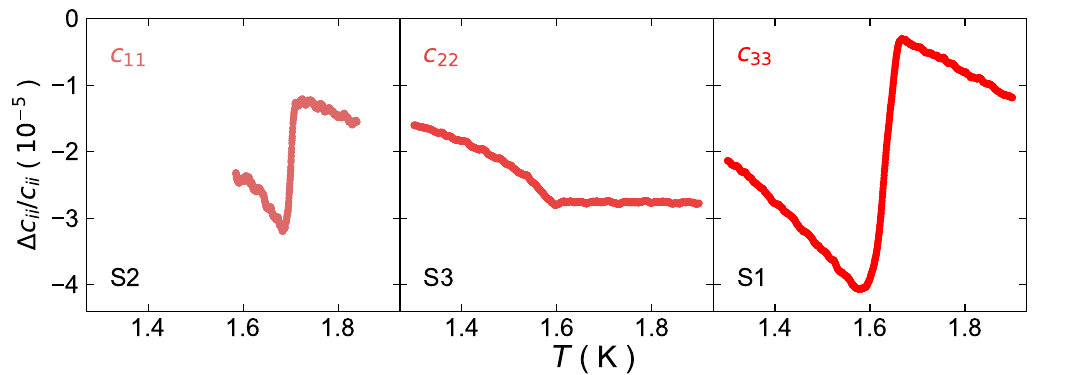}%
\caption{\textbf{Relative change in compressional elastic moduli through \Tc.} The compressional elastic moduli are shown as a function of temperature through \Tc. $c_{33}$ and $c_{11}$ were measured on a single-\Tc sample and $c_{22}$ was measured on the double-\Tc sample. Both $c_{11}$ and $c_{33}$ have clearly resolvable discontinuities at \Tc, whereas $c_{22}$ has a barely-resolvable discontinuity. $\Delta c_{ii}/c_{ii}$ is defined as in \autoref{fig:singleTc}.}%
\label{fig:compressionjumps}%
\end{figure}

We also measure the two other compressional moduli---$c_{11}$ and $c_{22}$---and show them along with $c_{33}$ in \autoref{fig:compressionjumps}. $c_{11}$ has a discontinuity of approximately 20 parts per million---roughly a factor of 2 smaller than the discontinuity in $c_{33}$. In contrast, $c_{22}$ has a discontinuity of at most 1 part per million---significantly smaller than the other two compressional moduli. Discontinuities in all three compressional moduli are allowed by symmetry for any superconducting order parameter (see \citet{ghosh_thermodynamic_2021} and the S.I.).

We first analyze the data using only the presence or absence of discontinuities in the elastic moduli. This analysis is based on symmetry arguments alone and is independent of the size of the discontinuities. We then perform a quantitative analysis of the discontinuities using Ehrenfest relations. Finally, we combine all of our observations to speculate on which particular superconducting order parameter is most consistent with our data.

\textbf{Symmetry of the superconducting order parameter.} The presence or absence of a discontinuity in each elastic modulus constrains the symmetry of the superconducting order parameter. Roughly speaking, only strains that couple linearly to a degree of freedom associated with the superconducting order parameter show discontinuities at \Tc. We illustrate this with a couple of examples; a more rigorous derivation is given in the SI. 

Discontinuities in elastic moduli arise when there is coupling between strain and superconductivity that is linear in strain and quadratic in the order parameter. For a single-component superconducting order parameter, this only occurs for compression strains \cite{rehwald1973study}. A single-component order parameter can be written as $\eta = \eta_0 e^{i \phi}$, where $\eta_0$ is the magnitude of the gap (which may depend on momentum) and $\phi$ is the superconducting phase. The lowest-order coupling to a strain $\epsilon_{ij}$ is $\epsilon_{ij}\eta^{\star}\eta = \epsilon_{ij}\eta_0^2$, where $(^{\star})$ denotes complex conjugation. This coupling is allowed only if $\epsilon_{ij}$ preserves the symmetry of the lattice, i.e. it is only allowed for compression strains and not for shear strains (which break the lattice symmetry). Since $\eta_0^2$ is proportional to the superfluid density, the physical interpretation of the resulting discontinuity at \Tc is that compression strain couples to the superfluid density, which turns on at \Tc and provides a new degree of freedom that softens the lattice. 

%The next order coupling, quadratic in both strain and order parameter (i.e. $\epsilon_{ij}^2 \eta^{\star}\eta$), is allowed for any order parameter and strain combination, but does not result in a discontinuity in the respective elastic modulus, only a change in the slope as a function of temperature \cite{rehwald1973study}.

In contrast to single-component order parameters, multi-component order parameters can have discontinuities in shear elastic moduli. This is because there are more degrees of freedom associated with a multi-component order parameter than with a single-component order parameter. Writing a two-component order parameter as $\vec{\eta} = \left\{\eta_{0,i} e^{i \phi_i},\eta_{0,j} e^{i \phi_j}\right\}$, there are now several possible couplings at lowest order. Taking the well-known $p$-wave state in tetragonal crystals as an example, one possible coupling is $\epsilon_{xy} \eta_{0,p_x} \eta_{0,p_y} \cos\left(\phi_x-\phi_y\right)$ \footnote{the fully symmetrized coupling is $1/4 \left(\epsilon_{xy} + \epsilon_{yx}\right)\left(\eta_{p_x}^{\star}\eta_{p_y}+\eta_{p_y}^{\star}\eta_{p_x}\right)$.}. This is the so-called ``phase mode'' of the order parameter, as it couples shear $xy$ strain to the relative phase of the two components (see \autoref{fig:schematic}). This produces a discontinuity in the associated elastic modulus $c_{66}$. The relative phase is a new degree of freedom that is only present in a multicomponent order parameter, as strain cannot couple to the absolute phase of a single-component order parameter (such a term would break gauge symmetry). Similar expressions exist for orthorhombic crystals (see SI for details), but the main conclusion is independent of the crystal structure: shear elastic moduli \textit{only} exhibit discontinuities at \Tc for multi-component superconducting order parameters.

The absence of a discontinuity in any shear elastic modulus in the single-transition samples (S1 and S2) rules out all uniform-parity \footnotemark[5]{}, two-component order parameters in \ute. While there are no natural two-component order parameters in \ute because the crystal structure is orthorhombic, many nearly or accidentally-degenerate order parameters have been proposed to explain the presence of the two nearly-degenerate \Tcs, time reversal symmetry breaking, and chiral surface states (see \autoref{tab:op}). One proposal is the onset of first a B$_{2u}$ state followed by a B$_{3u}$ state at the second, lower \Tc \cite{hayes2021multicomponent,weiInterplayMagnetismSuperconductivity2022,shishidou2021TopologicalBandSuperconductivity,choiCorrelatedNormalState2023,shafferChiralSuperconductivityMathrmUTe2022}. This proposal predicts the usual discontinuities in compressional moduli at the first (higher-temperature) \Tc, followed by a discontinuity in the compressional moduli \textit{and} the $c_{66}$ shear modulus at the lower \Tc. In fact, the product of any two odd-parity (i.e. $p$ or $f$-wave) states or any two even-parity states (i.e. $s$ or $d$-wave) in D$_{2h}$ predicts a discontinuity in either $c_{44}$, $c_{55}$, or $c_{66}$, none of which we observe. This strongly constrains the superconducting order parameter of \ute to be of the single-component type. Finally, we note that our data is fully consistent with \textit{any} single-component order parameter, including even-parity states like $s$-wave and $d$-wave.\footnotetext[5]{It is possible to come up with mixed-parity order parameters that do not have jumps in shear elastic moduli, e.g. $s + i p_x$. Roughly speaking, these mixed-parity order parameters do not produce jumps in shear moduli because the product of an even-parity and an odd-parity object is itself an odd-parity object. Strains, on the other hand, are always even-parity objects. Thus, a term that is one power of an even-parity order parameter, one power of an odd-parity order parameter, and one power of strain, is an odd-parity term. Such terms are forbidden, as the free energy is even parity. These scenarios have no experimental backing, but have been proposed on theoretical grounds \cite{ishizuka2021periodic}.}

The similar absence of discontinuities in the shear elastic moduli of the two-transition sample (S3) rules out the multi-component explanation for the second superconducting transition. We find that the single discontinuity in $c_{33}$ in single-\Tc samples is approximately the same size as the sum of the two discontinuities found in double-\Tc samples. This suggests that, below the second transition, all electrons in \ute are in the same thermodynamic state, rather than double-\Tc samples having two separate superconducting mechanisms. This suggests a common origin for the two superconducting transitions, perhaps split by local strains \cite{thomasSpatiallyInhomogeneousSuperconductivity2021a} or magnetic impurities \cite{sundarUbiquitousSpinFreezing2023}. Why this usually manifests as only two sharp \Tcs (as we also observe in our data), rather than multiple \Tcs or a broad transition, remains an open question. It also leaves unresolved the issue of why even single-\Tc samples become double-\Tc samples under hydrostatic pressure, leaving open the possibility of the appearance of a multi-component order parameter under pressure.

\begin{figure}%
\includegraphics[width=1\columnwidth]{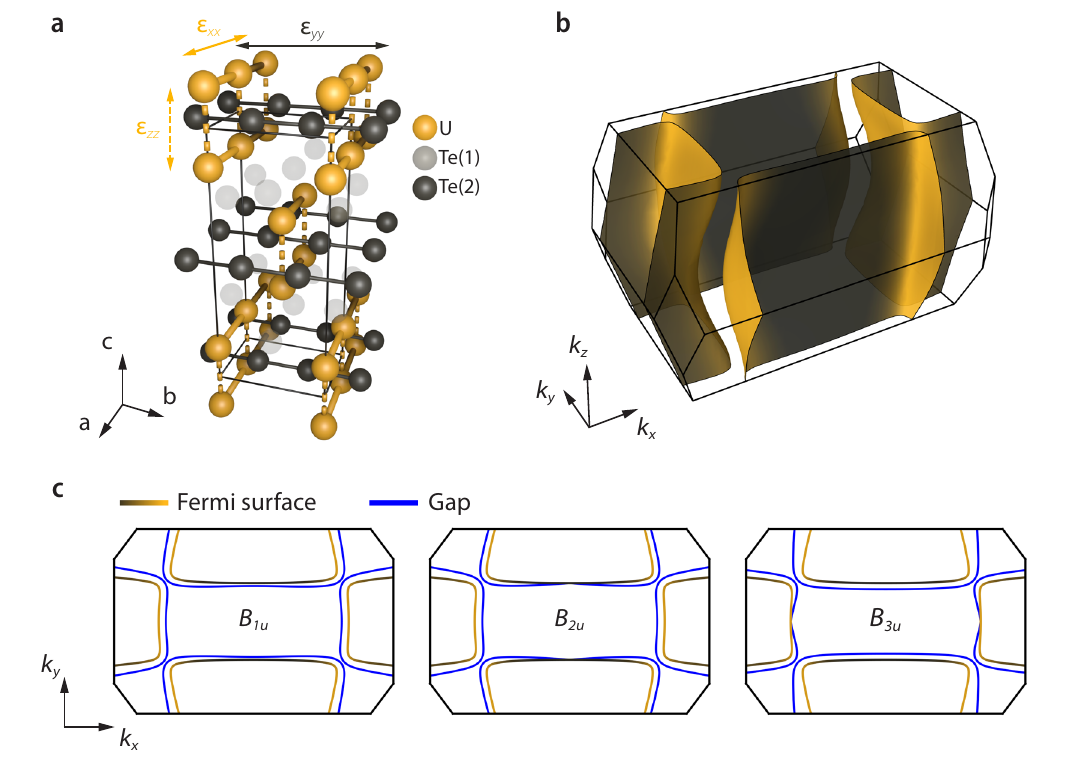}%
\caption{\textbf{Influence of compression strains on the crystal structure and Fermi surface of \ute.} Panel (a) shows the crystal structure of \ute. Highlighted are tellurium chains along the $b$ axis, and chains that run along the $a$ axis consisting of $c$-axis-oriented uranium dimers. These chains dominate the geometry of the Fermi surface shown in panel (b), modeled after quantum oscillation measurements \cite{eaton2023Quasi2DFermiSurface}. The Fermi surface is colored according to its uranium (yellow) and tellurium (gray) content. The superconducting gaps for three possible odd-parity order parameters are shown at $k_z=0$ as blue lines in panel (c).}%
\label{fig:crystal}%
\end{figure}

\textbf{Ehrenfest analysis and the coupling of compression strains to superconductivity.} The smallness of the discontinuity in $c_{22}$ compared to the other two compressional moduli indicates that the superconductivity in \ute is insensitive to strain along the $b$ axis ($\epsilon_{yy}$). This observation is made quantitative through Ehrenfest relations, which relate discontinuities in the elastic moduli, $\delta c_{ij}$, to the discontinuity in the specific heat, $\Delta C$. The Ehrenfest relations are
\begin{equation}
\delta c_{ij} = - \frac{\Delta C}{T}\left(\frac{d\Tc}{d \epsilon_{ij}}\right)^2,
\label{eq:ehrenfest}
\end{equation} 
where $\frac{d\Tc}{d \epsilon_{ij}}$ is the derivative taken at zero applied stress. Using the specific heat measured on sample S3 (see S.I.) and the data shown in \autoref{fig:compressionjumps}, we calculate $\frac{d\Tc}{d \epsilon_{xx}} = 0.23 \pm 0.02$ K/(\% strain), $\frac{d\Tc}{d \epsilon_{yy}} = 0.07 \pm 0.02$ K/(\% strain), and $\frac{d\Tc}{d \epsilon_{zz}} = 0.34 \pm 0.02$ K/(\% strain). These values are roughly consistent with those measured in uniaxial strain experiments \cite{girod2022ThermodynamicElectricalTransport} (see S.I. for the quantitative comparison).

These Ehrenfest relations indicate that the superconductivity of \ute is significantly more sensitive to strains along the $a$ and $c$ axes than it is to strain along the $b$ axis. This observation is perhaps surprising given the relatively quasi-two-dimensional nature of the Fermi surface measured by quantum oscillations in \ute \cite{eaton2023Quasi2DFermiSurface,aoki2022FirstObservationHaas}. The Fermi surface consists of two sets of quasi-one-dimensional sheets running along the $a$ and $b$ axes that hybridize to form one electron and one hole pocket (see \autoref{fig:crystal}). Thus, if any direction is to be weakly coupled to superconductivity, one might expect it to be the $c$ axis. This argument is unchanged by the possible existence of an additional small pocket of Fermi surface reported by \citet{broyles2023revealing}, as it is roughly isotropic and thus does not single out any particular direction.

Looking at the crystal structure in \autoref{fig:crystal}, however, it is clear that the $a$ and $b$ axes are highly asymmetric: chains of $c$-axis-coupled uranium dimers run along the $a$ axis, whereas chains of tellurium run along the $b$ axis (the other tellurium site, Te(1), participates much less in the Fermi surface than the Te(2) chains: see SI.). Thus $\epsilon_{xx}$ and $\epsilon_{zz}$ modulate the inter and intra-dimer coupling of the uranium dimers, respectively, whereas $\epsilon_{yy}$ only modulates the weak inter-chain coupling of the uranium chains. $\epsilon_{yy}$ does, however, modulate the intra-chain spacing of tellurium chains along the $b$ axis. Our observation of the relative insensitivity of \Tc to $\epsilon_{yy}$ therefore suggests that the superconducting pairing is more sensitive to the uranium-uranium distances than to the tellurium-tellurium distances.

\textbf{Proposed single-component superconducting order parameter.} Thermal transport \cite{metz2019PointnodeGapStructure}, specific heat \cite{ranNearlyFerromagneticSpintriplet2019,kittaka2020OrientationPointNodes}, and penetration depth \cite{ishiharaChiralSuperconductivityUTe22023} all suggest the presence of point nodes in the superconducting gap of \ute. $B_{1u}$, $B_{2u}$, and $B_{3u}$ order parameters all have point nodes in their superconducting gaps, but these nodes lie along different directions in momentum space and thus intersect different portions of the Fermi surface (or may not intersect the Fermi surface at all if it is quasi-2D). 

We use our observation of relatively weak coupling between $
\epsilon_{yy}$ and \Tc to motivate a particular orientation of the point nodes in \ute and to suggest one particular single-component order parameter. \autoref{fig:crystal} shows a tight binding model of the Fermi surface of \ute as determined by quantum oscillations, color-coded by the relative uranium 6$d$ and tellurium 5$p$ content (both bands have significant uranium 5$f$ character that contribute to their heavy masses, but not to their geometry). Our results suggest that the superconducting gap is either weak or absent on the tellurium-dominant electron Fermi surface. Only the $B_{2u}$ order parameter has nodes that lie along the $k_y$ direction, producing a node in the gap on the tellurium-dominant surface and a gap maximum on the uranium-dominant surface.  We note that a reported small pocket with a light mass does not qualitatively affect this argument \cite{broyles2023revealing}, as it is largely isotropic in shape and thus will not respond differently to $\epsilon_{xx}$ and $\epsilon_{yy}$ strain.

\section{Discussion}

Our primary result is that the superconducting transitions in both single and double-transition samples of \ute exhibit no thermodynamic discontinuities in any of the shear elastic moduli at \Tc. The strictest interpretation of this result is that it rules out all multi-component order parameters that have a bilinear coupling to strain. For \ute, this rules out all multi-component order parameter scenarios except for mixed-parity order parameters like $s$ + $p$ wave \footnotemark[5]{}.

Looking beyond our own experiment, there is strong evidence that \ute has an odd-parity order parameter. There is also strong evidence for nodes in the superconducting gap. Combined with our result, this leaves the three $B_{iu}$ representations as the only possibilities. We argue that our observation of a lack of sensitivity of \Tc to $\epsilon_{yy}$ suggests a $B_{2u}$ order parameter.

A single-component order parameter places constraints on possible explanations for other experiments. First, a single-component order parameter cannot break time reversal symmetry. This suggests that the interpretation of time reversal symmetry breaking at \Tc as seen by polar Kerr effect measurements \cite{hayes2021multicomponent,weiInterplayMagnetismSuperconductivity2022}, along with the chiral surface states seen in STM \cite{Jiao2020} and microwave surface impedance measurements \cite{baeAnomalousNormalFluid2021}, may need to be revisited.

The search for multi-component superconductors continues: they are of both fundamental and practical interest, since a multi-component order parameter is a straightforward route to topological superconductivity. We find that, while \ute may have an odd-parity, spin-triplet order parameter, it seems that the most likely order parameter to condense at \Tc is of the single-component $B_{2u}$ representation---either $p_y$ or $f_{yz^2}$-wave superconductivity. Definitive determination of the orientation of the nodes in the superconducting gap would confirm this scenario.

\newpage
\section{Methods}

\subsection{Sample Growth and Preparation}
Single crystals of \ute were grown by the chemical vapor transport (CVT) method as described in \citet{ranNearlyFerromagneticSpintriplet2019,ranComparisonTwoDifferent2021a}. Samples with one \Tc (two \Tcs) were grown in a two-zone tube furnace with temperatures of $950^{\circ}$C and $860^{\circ}$C ($1060^{\circ}$C and $1000^{\circ}$C) at the hot and cold end, respectively.

Specimens were aligned to better than $1^{\circ}$ using their magnetic anisotropy (performed in a Quantum Design MPMS) and X-ray diffraction (performed in a Laue backscattering system) measurements. Samples were then polished to produce two parallel faces normal to the $(100)$, $(010)$, and $(001)$ directions, depending on the mode geometry (see \autoref{table:pulse echo measurements}). 

% "We used x-ray diffraction (Rigaku Miniflex and Photonic Laue backscattering system), and magnetization anisotropy (MPMS) to determine to crystal direction of UTe2. Oriented crystals were mounted on a polishing chuck with a crystal bond (maximum baking temperature no higher than 150C). The desired sample shape and thickness were obtained using SiC polishing pads (Buehler Microcut P2500 and P4000) and isopropanol as a lubricant. After thoroughly cleaning the sample and the polishing chuck, the sample surface was then fine-polished on a polishing cloth (Buehler TexMet C) and a slurry that consist of 0.3-micron aluminum oxide (Buehler Micropolish) in isopropanol.”

Thin-film ZnO piezoelectric transducers were sputtered from a ZnO target in an atmosphere of oxygen and argon. Both shear and longitudinal responses are present in each transducer---the shear axis was aligned with either the $(100)$, $(010)$, or $(001)$, again depending on the particular mode geometry. 3 crystals were measured in total; see \autoref{table:pulse echo measurements} for details.
The shear response in our deposited transducers was achieved by mounting the sample on the far end of the sputtering sample stage, maximizing the distance between sample and ZnO target. The position of the sample stage was fixed during the entire deposition process (i.e. rotation was disabled on the sample stage). The resulting polarization direction of the generated sound wave is then parallel to the shortest line drawn between the sample and the target---this orientation was verified using the absolute speed of sound and the moduli obtained using resonant ultrasound spectroscopy \cite{theuss2023resonant}.
\begin{table}[]
	% \centering
	\begin{tabular}{cccccccc}
		\toprule
		\# $T_c$           & Sample              & $\vec{k}$                & $\vec{u}$ & $c_{ij}$ & $f$ (MHz) & $d$ ($\mu$m) & $c$ (GPa)\\
        \midrule
        \multirow{5}{*}{1} & \multirow{3}{*}{S1} & \multirow{3}{*}{$[001]$} & $[100]$   & $c_{55}$ & 1261      & $330\pm17$ & $51\pm5$\\
                           &                     &                          & $[010]$   & $c_{44}$ & 1434      & $330\pm17$ & $27\pm3$\\
                           &                     &                          & $[001]$   & $c_{33}$ & 2260      & $330\pm17$ & $91\pm11$\\
                           \cmidrule{2-8}
                           & \multirow{2}{*}{S2} & \multirow{2}{*}{$[100]$} & $[100]$   & $c_{11}$ & 823       & $920\pm46$ & $81\pm8$\\
                           &                     &                          & $[010]$   & $c_{66}$ & 1250      & $920\pm46$ & $28\pm3$\\
        \midrule
        \multirow{5}{*}{2} & \multirow{5}{*}{S3} & \multirow{3}{*}{$[001]$} & $[100]$   & $c_{55}$ & 1348      & $550\pm28$ & $52\pm5$\\
                           &                     &                          & $[010]$   & $c_{44}$ & 1352      & $550\pm28$ & $28\pm3$\\
                           &                     &                          & $[001]$   & $c_{33}$ & 1348      & $550\pm28$ & $88\pm9$\\
                           \cmidrule{3-8}
                           &                     & \multirow{2}{*}{$[010]$} & $[100]$   & $c_{66}$ & 1362      & $290\pm15$ & $30\pm3$\\
                           &                     &                          & $[010]$   & $c_{22}$ & 1362      & $290\pm15$ & $141\pm15$\\
        \bottomrule
	\end{tabular}
	\caption{\textbf{Sample configuration.} Listed are the transducer configurations for all the measurements in this manuscript. Samples are sorted by the number of superconducting phase transitions (first column). Additional information given is the propagation $\vec{k}$ and the polarization $\vec{u}$ of the sound pulse excited in the sample, as well as the measured elastic modulus. Also shown is the frequency at which each measurement is performed. We also provide the thicknesses $d$ of the measured samples and the resulting absolute values of the elastic moduli obtained from the separation of echoes at room temperature. Uncertainties represent a 5~\% uncertainty on the thickness.}
	\label{table:pulse echo measurements}
\end{table}

\subsection{Pulse-Echo Ultrasound Measurements}
Measurements were performed in an Oxford Instruments Heliox $^3$He refrigerator. We used a traditional phase-comparison pulse-echo ultrasound method to measure the change in elastic moduli relative to the highest temperature $T_0$, i.e. we measured $\Delta c/c \equiv \left(c(T)-c(T_0)\right)/c(T_0)$. Short bursts (typically $\sim 50$ ns) of radiofrequency signals, with the carrier frequency between 500 MHz and 2.5 GHz, were generated with a Tektronix TSG 4106A RF generator modulated by a Tektronix AFG 31052 arbitrary function generator, amplified by a Mini-Circuits ZHL-42W+ power amplifier, and transmitted to the transducer. The signal was detected with the same transducer, amplified with a Mini-Circuits ZX60-3018G-S+ amplifier, and recorded on a Tektronix MSO64 oscilloscope. The detection amplifier was isolated from the power amplifier using Mini-Circuits ZFSWA2-63DR+ switches, timed with the same Tektronix AFG 31052 arbitrary function generator.

\begin{figure}[h]%
    \includegraphics[width=1\columnwidth]{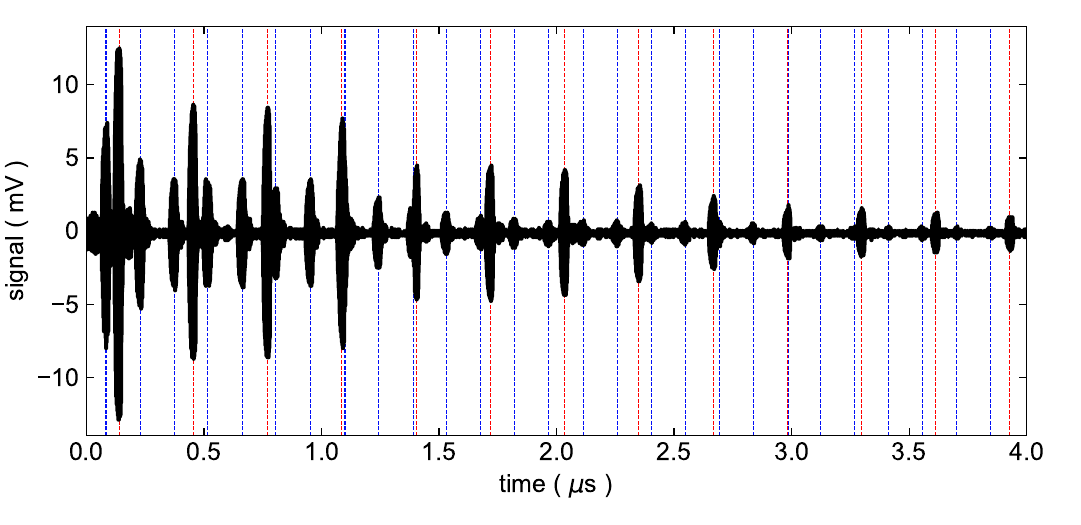}%
    \caption{\textbf{Raw Pulse-Echo Signal.} The raw signal from a sputtered ZnO shear transducer on sample S3 with sound propagation along the [010] and polarization along the [100] directions. The transducer exhibits both a compressional (blue lines) and a shear (red dashed lines) response. These correspond to sound modes determined by the elastic moduli $c_{22}$ and $c_{66}$, respectively.}%
    \label{fig:raw pulse train}%
\end{figure}

Both shear and compressional sound are generated by our transducers---these signals are separated in the time domain due to the different speeds of propagation and identified as shear or compression using the known elastic moduli of \ute \cite{theuss2023resonant}. \autoref{fig:raw pulse train} shows a raw pulse-echo signal from a transducer sputtered on sample S3 with sound propagating along the [010] direction with a shear polarization axis along [100], thus measuring $c_{22}$ and $c_{66}$ simultaneously. Echoes corresponding to the different elastic modes can be clearly identified as shear (red vertical dashed lines) and compression (blue vertical dashed lines).

\begin{figure}[h]%
    \includegraphics[width=1\columnwidth]{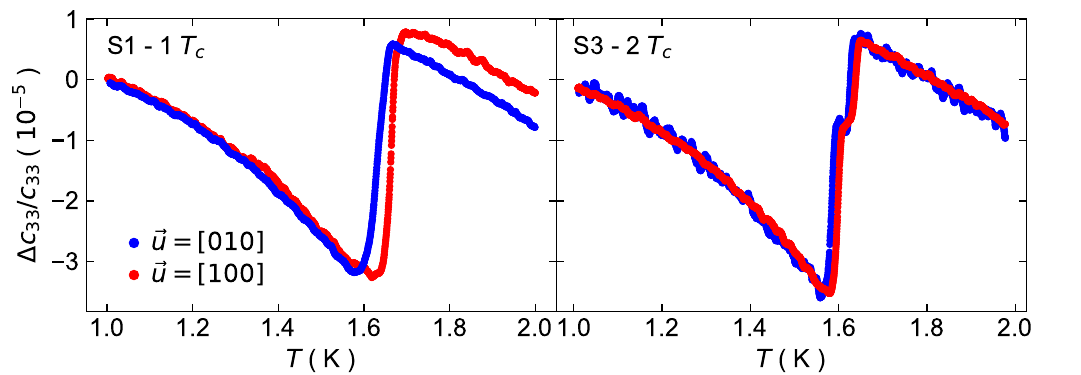}%
    \caption{\textbf{Transducer Comparison.} Shown are $\Delta c_{33}/c_{33}$ for single \Tc (S1, left) and two \Tc (S3, right) samples. For each sample we compare the relative change in elastic modulus between measurements obtained with two different transducers. Both transducers excited sound along the [001] direction. However, for the data in red, the shear component of the transducer was polarized along [100] (additionally measuring $c_{55}$), whereas for the data in blue, the shear component of the transducer was polarized along [010] (additionally measuring $c_{44}$).}%
    \label{fig:transducer comparison}%
\end{figure}

The phase of each echo was analyzed using a software lockin, and the relative change in phase between two echoes was converted to the relative change in speed of sound as a function of temperature. In \autoref{fig:transducer comparison} we compare the temperature dependence of $c_{33}$ of samples S1 and S3 obtained with different transducers.

\section{Data Availability}
Data that support the plots within this paper and other findings of this study are available from the corresponding author upon reasonable request.

\section{Acknowledgments}
We acknowledge helpful discussions with D. Agterberg and P. Brydon. B.J.R. and F.T. acknowledge funding from the Office of Basic Energy Sciences of the United States Department of Energy under award no. DE-SC0020143 for preparing the samples and transducers, performing the measurements, analyzing the data, and writing the manuscript. Research at the University of Maryland was supported by the Department of Energy award number DE-SC-0019154 (sample characterization), the Gordon and Betty Moore Foundation’s EPiQS Initiative through grant number GBMF9071 (materials synthesis), the National Science Foundation under grant number DMR-2105191 (sample preparation), the Maryland Quantum Materials Center and the National Institute of Standards and Technology. A part of this work was performed at the Cornell Center for Materials Research Shared Facilities which are supported through the NSF MRSEC program (DMR-1719875).

%%%%%%%%%%%%% input SI
% \newpage
\clearpage

\section{SUPPLEMENTARY INFORMATION}
\subsection{Data Reproducibility}
\begin{figure}[H]%
\includegraphics[width=1\columnwidth]{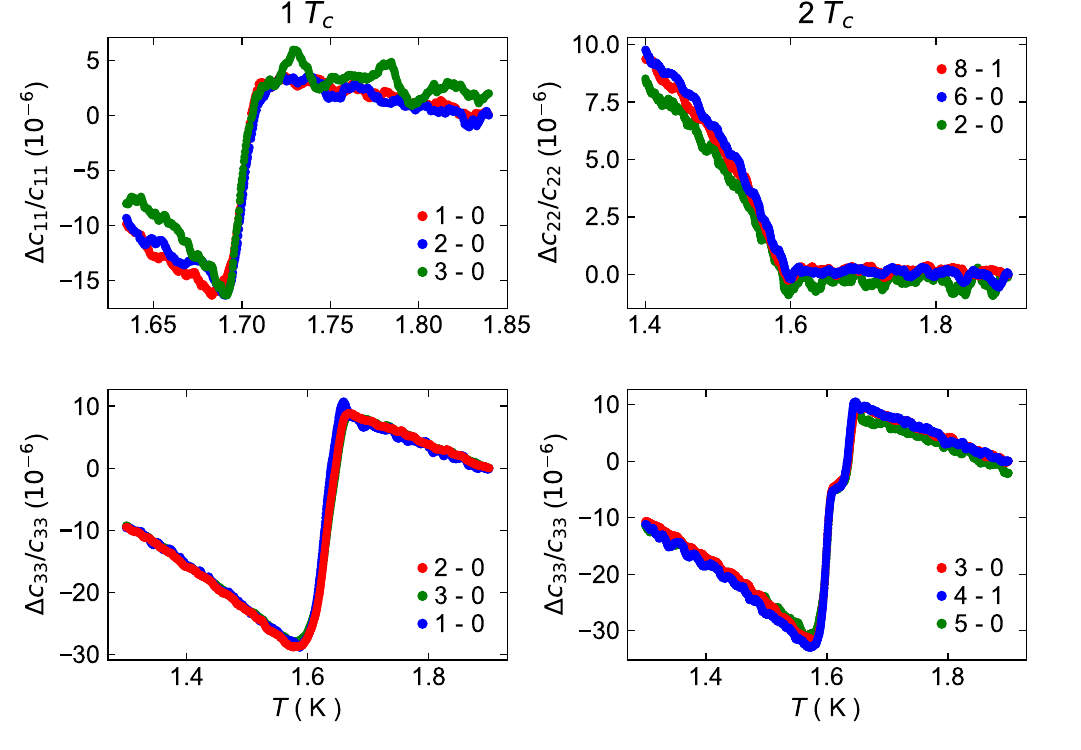}%
\caption{\textbf{Echo Analysis Compressional Moduli.} The relative change of compressional elastic moduli as obtained from different echoes in a single pulse-echo ultrasound experiment. The colors indicate the echoes used for each curve. The data in red are the data shown in the main. The left column shows data for samples with one superconducting transition, the right column is for samples with two transitions.}%
\label{fig:different echoes compression}%
\end{figure}

\begin{figure}[H]%
\includegraphics[width=1\columnwidth]{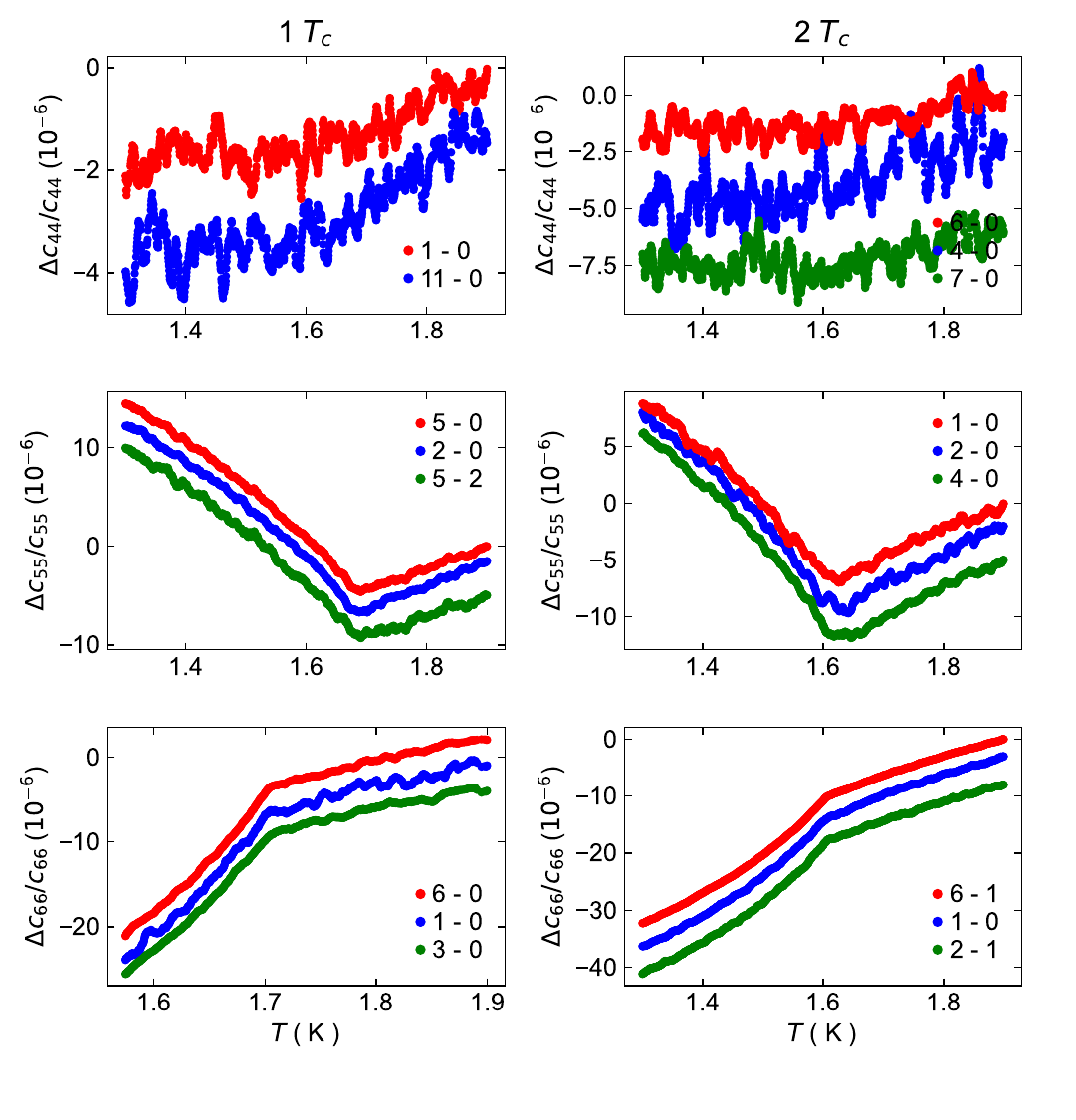}%
\caption{\textbf{Echo Analysis Shear Moduli.} The relative change of shear elastic moduli as obtained from different echoes in a single pulse-echo ultrasound experiment. The colors indicate the echoes used for each curve. The data in red are the data shown in the main. The left column shows data for samples with one superconducting transition, the right column is for samples with two transitions.}
\label{fig:different echoes shear}%
\end{figure}

\begin{figure}[H]%
\includegraphics[width=1\columnwidth]{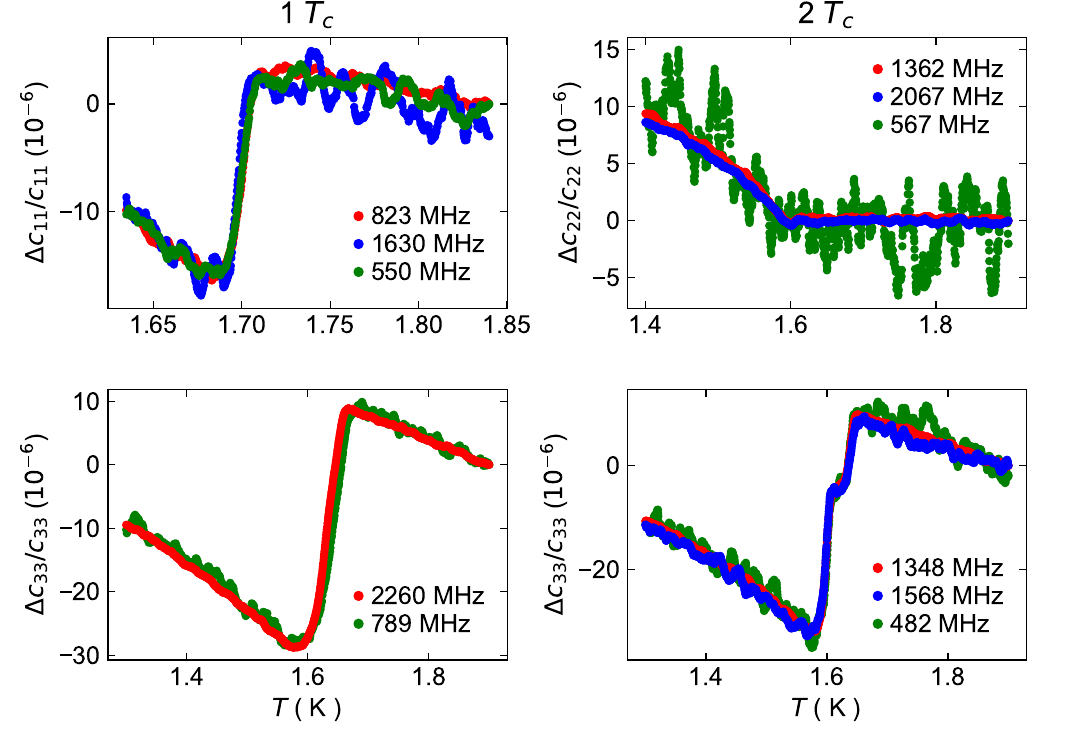}%
\caption{\textbf{Frequency Dependence Compressional Moduli.} The relative change of compressional elastic moduli at different frequencies. The data in red are the data shown in the main. The left column shows data for samples with one superconducting transition, the right column is for samples with two transitions.}
\label{fig:frequency dependence compression}%
\end{figure}

\begin{figure}[H]
\includegraphics[width=1\columnwidth]{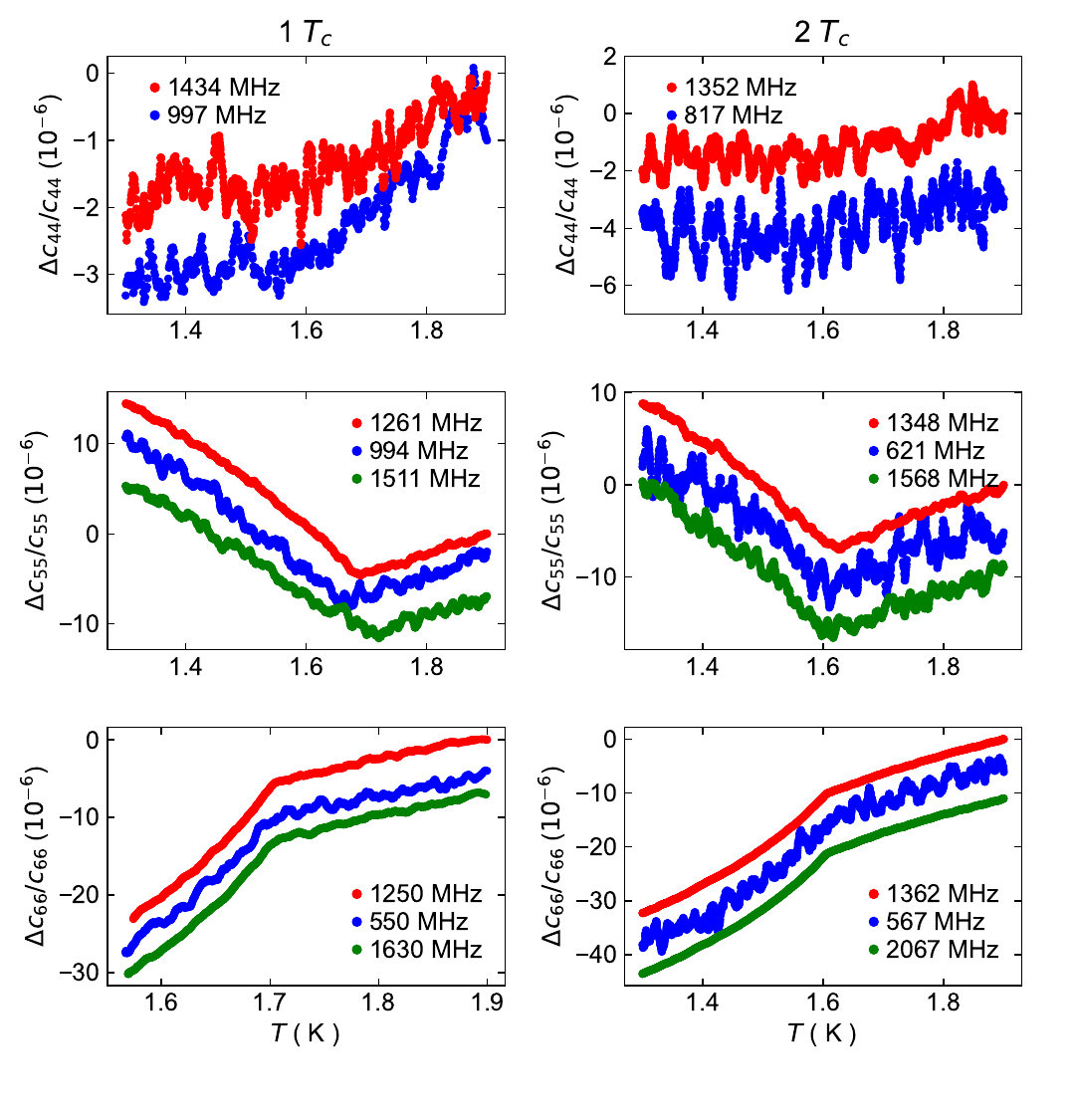}%
\caption{\textbf{Frequency Dependence Shear Moduli.} The relative change of shear elastic moduli at different frequencies. The data in red are the data shown in the main. The left column shows data for samples with one superconducting transition, the right column is for samples with two transitions.}
\label{fig:frequency dependence shear}%
\end{figure}

\autoref{fig:different echoes compression} and \autoref{fig:different echoes shear} show the relative change of elastic moduli as a function of temperature as obtained when different echoes from a single experiment are used for the data analysis. \autoref{fig:frequency dependence compression} and \autoref{fig:frequency dependence shear} show the relative change of elastic moduli for different carrier frequencies of the excited sound pulse. We find no significant dependence on either the echoes, or the frequencies used for any of our measurements.

\pagebreak
\subsection{Noise Analysis}
\begin{figure}[H]%
\includegraphics[width=1\columnwidth]{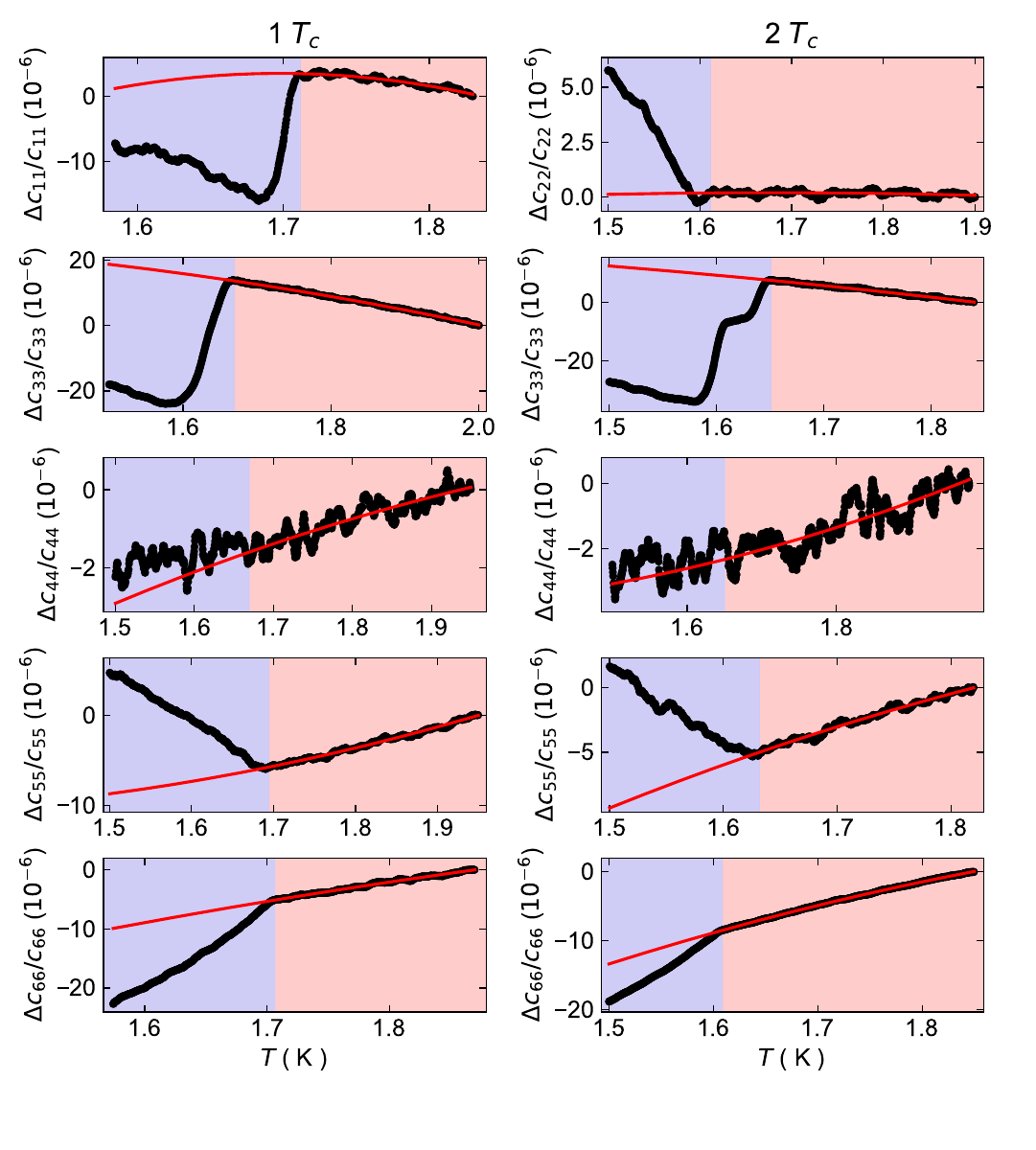}%
\caption{\textbf{Background Subtraction.} Temperature dependence of all elastic moduli shown in the main (black points). A second order polynomial (red line) has been fitted to each modulus in the normal state above $T_c$ (red shaded region). The left column shows data for samples with one superconducting transition, the right column is for samples with two transitions.}%
\label{fig:noise estimate background}%
\end{figure}

\begin{figure}[H]%
\includegraphics[width=1\columnwidth]{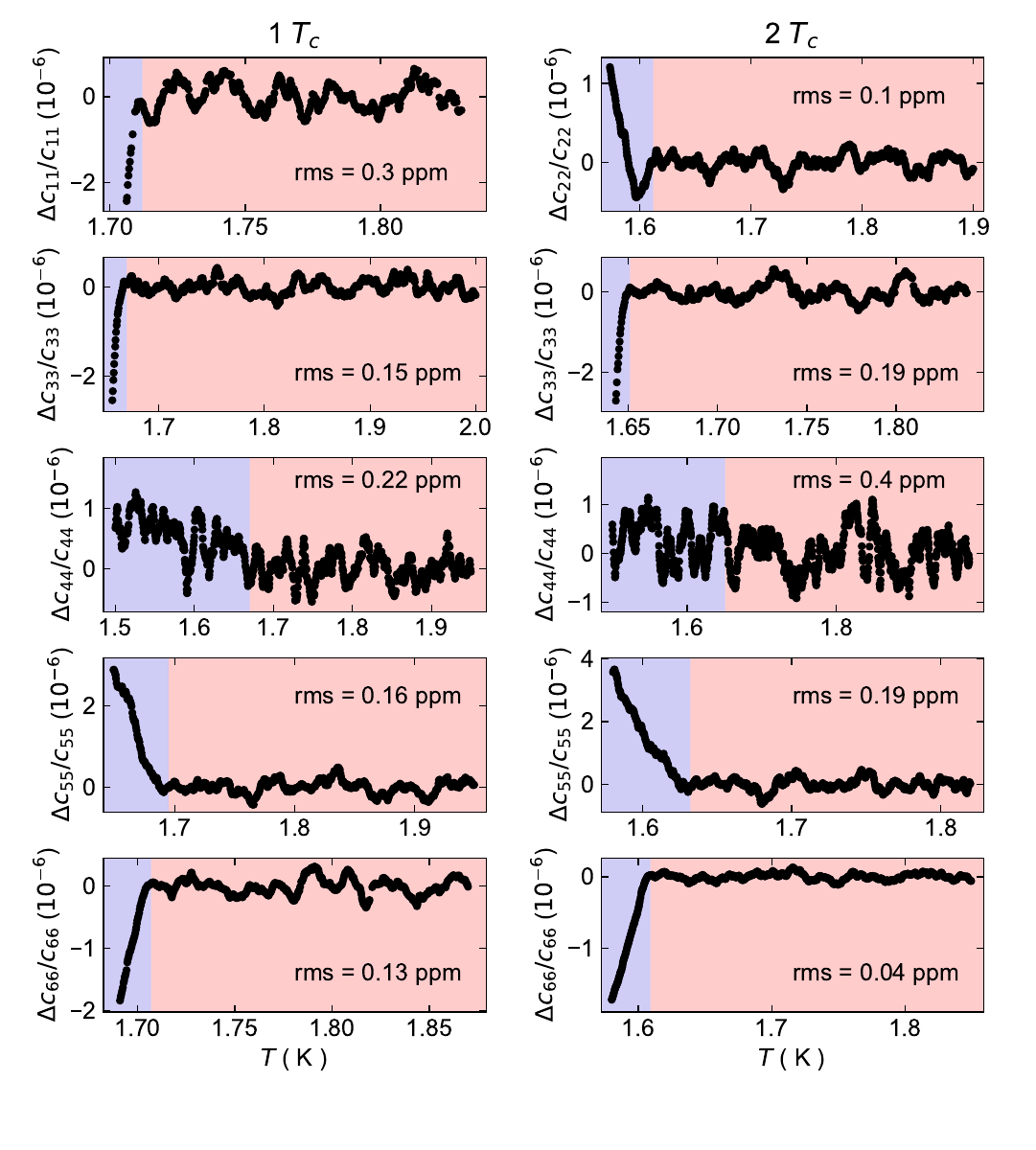}%
\caption{\textbf{Noise Estimate.} Elastic moduli from \autoref{fig:noise estimate background} with the normal state background (see \autoref{fig:noise estimate background}) subtracted. A RMS is calculated for each modulus in the same temperature range that was used to fit the background in \autoref{fig:noise estimate background} (red shaded region). The purple/red shaded temperature regions are identical to those in \autoref{fig:noise estimate background}. The left column shows data for samples with one superconducting transition, the right column is for samples with two transitions.}%
\label{fig:noise estimate final}%
\end{figure}

\autoref{fig:noise estimate background} shows the relative change of all elastic moduli also shown in the main. In order to estimate the noise of our data, a second order polynomial has been fitted to the normal state data (highlighted by a red background in \autoref{fig:noise estimate background}). In \autoref{fig:noise estimate final} we show the same elastic moduli with that polynomial subtracted from the data. We then estimate its noise as the RMS of the background-subtracted data above the transition, i.e. the same temperature range which we used to fit said polynomial background (red shaded region). The resulting RMS values lie between 0.04~ppm and 0.41~ppm (on average less than $1.9\times 10^{-7}$).
\newpage

\pagebreak
\subsection{Landau Free Energy Calculations}
Elastic moduli are the second derivative of the free energy with respect to strain, i.e. they are the strain susceptibility, in analogy with the heat capacity, which is the second derivative of the free energy with respect to temperature. If strain couples linearly to the square of the order parameter $\eta$ (just like temperature does in the term $(T-T_c) \eta^2$), the respective elastic modulus will exhibit a discontinuity at the phase transition (just like the specific heat does). The reason for these discontinuities is that immediately below $T_c$, the system has a new degree of freedom that can respond when you apply strain (in the case of elastic moduli) or change temperature (in the case of heat capacity). This new degree of freedom means that the response below $T_c$ is entirely distinct from that above $T_c$: even though the order parameter itself changes continuously, the system's susceptibility to changes in the order parameter is discontinuous.

For a single-component order parameter, only compressional moduli can exhibit this discontinuity. For a two-component order parameter, on the other hand, discontinuities in compressional moduli \textit{and} certain shear moduli are allowed. This is because for single-component order parameters, the only quantity that can respond to strain is the magnitude of the order parameter. The bare “amplitude” of the order parameter breaks gauge symmetry and thus cannot enter directly into the free energy or couple linearly to external parameters like strain. As the magnitude of the order parameter is a scalar (simply a number), this means that it couples to scalar strains, i.e. compressional moduli.

For a two component order parameter, there are two new gauge-invariant quantities that can couple to strain: the relative phase between the components of the order parameter, and the overall “orientation” of the two components in order-parameter space. These are new degrees of freedom that can be probed by shear strain, and thus are what allow for discontinuities in the shear moduli at $T_c$.

Below, we elaborate on these concepts within the Landau theory of second-order phase transitions.

\subsubsection{Elastic Free Energy}
The elastic free energy of a solid is given by $\frac{1}{2} \sum_{i,j} \varepsilon_i c_{ij} \varepsilon_j$, with strain $\vec{\varepsilon} = \{\varepsilon_{xx},\varepsilon_{yy},\varepsilon_{zz},2\varepsilon_{yz},2\varepsilon_{xz},2\varepsilon_{xy}\}$ and the elastic tensor $\boldsymbol{c}$ in Voigt notation. In an orthorhombic crystal environment (i.e. point group $D_{2h}$), all individual elements of the strain tensor transform as a particular irreducible representation of the point group $D_{2h}$. In particular, we can rewrite
\begin{equation}
    \label{eq:irreducible strains}
    \vec{\varepsilon} = \{\varepsilon_{xx},\varepsilon_{yy},\varepsilon_{zz},2\varepsilon_{yz},2\varepsilon_{xz},2\varepsilon_{xy}\} = \{\varepsilon_{Ag,1},\varepsilon_{Ag,2},\varepsilon_{Ag,3},\varepsilon_{B3g},\varepsilon_{B2g},\varepsilon_{B1g}\},
\end{equation}
where the subscript now refers to the irreducible representation. Consequently, the elastic free energy can be rewritten as
\begin{align}
    \label{eq:elastic free energy}
    f_{el} = &\frac{1}{2} \left( c_{Ag,1} \varepsilon_{Ag,1}^2 + c_{Ag,2} \varepsilon_{Ag,2}^2 + c_{Ag,3} \varepsilon_{Ag,3}^2  + 
    2 c_{Ag,4} \varepsilon_{Ag,1} \varepsilon_{Ag,2} + 2 c_{Ag,5} \varepsilon_{Ag,1} \varepsilon_{Ag,3} + 2 c_{Ag,6} \varepsilon_{Ag,2} \varepsilon_{Ag,3}\right.\\
    \notag
    &\left. + c_{B3g} \varepsilon_{B3g}^2 + c_{B2g} \varepsilon_{B2g}^2 + c_{B2g} \varepsilon_{B2g}^2 \right).
\end{align}
Here, we have rewritten the elastic tensor according to
\begin{equation}
    \label{eq:elastic tensor}
	\boldsymbol{c} = 
	\begin{pmatrix}
	c_{11} & c_{12} & c_{13} &  0     &     0  &  0     \\
	c_{12} & c_{22} & c_{23} &  0     &     0  &  0     \\
	c_{13} & c_{23} & c_{33} &  0     &     0  &  0     \\
	0      & 0      & 0      & c_{44} &     0  &  0     \\
	0      & 0      & 0      &  0     & c_{55} &  0     \\
	0      & 0      & 0      &  0     &     0  &  c_{66}\\  
	\end{pmatrix} =
    \begin{pmatrix}
        c_{Ag,1} & c_{Ag,4} & c_{Ag,5} &  0      &     0   &  0      \\
        c_{Ag,4} & c_{Ag,2} & c_{Ag,6} &  0      &     0   &  0      \\
        c_{Ag,5} & c_{Ag,6} & c_{Ag,3} &  0      &     0   &  0      \\
        0        & 0        & 0        & c_{B3g} &     0   &  0      \\
        0        & 0        & 0        &  0      & c_{B2g} &  0      \\
        0        & 0        & 0        &  0      &     0   &  c_{B1g}\\  
        \end{pmatrix}.
\end{equation}
% Labeling the elastic moduli and corresponding strains according to their irreducible representation will make the analysis of strain-to-order-parameter coupling in the following sections straightforward.

\subsubsection{Order Parameter Free Energy and Coupling to Strain:\\One-Component Order Parameter}
A single-component superconducting order parameter (OP) can be parametrized as $\eta e^{i \gamma}$, with an amplitude $\eta$ and phase $\gamma$, both real. However, since the free energy needs to obey global gauge symmetry, the OP can only appear in even powers and the phase factor $e^{i \gamma}$ becomes unobservable. Only one degree of freedom remains, the amplitude (or superfluid density) $\eta$. The phase factor is thus dropped in the following discussion. In this case, the OP free energy expansion to fourth order reads
\begin{equation}
    \label{eq:OP free energy 1 component}
    f_{OP} = \frac{a}{2} \eta^2 + \frac{b}{4} \eta^4,
\end{equation}
where $a=a_0 \left( T - T_c \right)$. $a_0 >0$, and $b$ are phenomenological constants, $T_c$ is the critical temperature.

Since the OP has to appear in even powers in the free energy, the lowest possible coupling to strain is quadratic in OP and linear in strain. Furthermore, since the OP transforms as a one-dimensional irreducible representation of $D_{2h}$, its bilinear will always transform as the $A_g$ irreducible representation, irrespective of the particular representation. Thus, quadratic in OP and linear in strain coupling terms are only allowed for $A_g$ strains, and to lowest order, the terms in the free energy coupling the OP to strain are
\begin{equation}
    \label{eq:coupling free energy 1 component}
    f_{coupling} = \frac{1}{2} \left( g_1 \varepsilon_{Ag,1} + g_2 \varepsilon_{Ag,2} + g_3 \varepsilon_{Ag,3} \right) \eta^2.
\end{equation}
Following the formalism outlined in \cite{ghosh_thermodynamic_2021}, coupling of strain to the OP leads to a discontinuity of the respective elastic moduli at $T_c$ according to
\begin{equation}
    \label{eq:jump in el mod for one component OP}
    \delta c_{mn} = \left. - \frac{Z_m Z_n}{Y} \right|_{\eta\rightarrow\eta_0,\varepsilon_m \rightarrow 0},
\end{equation}
where $Z_m = \frac{\partial^2 f_{coupling}}{\partial \eta \partial \varepsilon_m} $, $Y = \frac{\partial^2 f_{OP}}{\partial \eta^2}$, and $\eta_0 = \sqrt{\frac{-a}{b}}$ is the equilibrium value for the OP defined by $\partial f_{OP}/\partial \eta = 0$. From \autoref{eq:jump in el mod for one component OP} it is straightforward to see that coupling terms in the free energy which are quadratic or higher order in strain will not lead to a discontinuity of the respective elastic modulus at $T_c$, which justifies the truncation of \autoref{eq:coupling free energy 1 component} after terms linear in strain. Consequently, in the case of a one-component OP, no shear modulus (i.e. $c_{B1g}$, $c_{B2g}$, or $c_{B3g}$) is allowed to show a discontinuity at $T_c$ (note that a discontinuity in its derivative is allowed \cite{Ramshaw2015a}). This is a general statement purely based on the dimensionality of the order parameter and irrespective of its particular irreducible representation. Combining \autoref{eq:coupling free energy 1 component} and \autoref{eq:jump in el mod for one component OP}, all $A_g$ elastic moduli exhibit a discontinuity at the critical temperature. The magnitudes of these discontinuities based on the free energy in \autoref{eq:OP free energy 1 component} and \autoref{eq:coupling free energy 1 component} are summarized in \autoref{table:elastic moduli jumps for different order parameters}.
\begin{table}[]
	% \centering
	\begin{tabular}{ccc}
		\toprule
		                   & One-Component OP                        & $B_{2u}+iB_{3u}$ \\
        % OP                 & $\eta$                                  & $\boldsymbol{\eta}=\eta \left(\cos \theta, e^{i\gamma} \sin \theta\right)$\\
                           & $\eta_0 = \sqrt{\frac{-a}{b}}$          & $\left(\theta_0,\gamma_0\right)=\left(\pi/4,\pm\pi/2 \right)$\\
                           &                                         & $\eta_0 = \pm \sqrt{\frac{-2 a_1}{b_1+b_3-b_4}}$\\
		\midrule
		$\delta c_{Ag,i}$ & $-\frac{g_i^2}{2 b}   \left(\eta\right)$  & $\frac{a_1 (-2 a_2 ( b_1 + b_3 - b_4 ) g_i^a g_i^s + a_1 ( - ( b_1+ b_3 - b_4 ) (g_i^a)^2 + 4 b_2 g_i^a g_i^s - ( b_1 - b_3 + b_4 ) (g_i^s)^2))}{a_1^2 (b_1^2 - 4 b_2^2 - (b_3 - b_4)^2) + 4 a_1 a_2 b_2 (b_1 + b_3 - b_4) - a_2^2 (b_1 + b_3 - b_4)^2} \left(\eta, \theta\right)$ \\
        $\delta c_{B1g}$   & 0                                       & $-\frac{g_4^2}{2 b_4} \left(\gamma\right)$  \\
        $\delta c_{B2g}$   & 0                                       &    0              \\
        $\delta c_{B3g}$   & 0                                       &    0              \\
        \bottomrule
	\end{tabular}
	\caption{\textbf{Discontinuities in elastic moduli for different OP configurations.} Magnitudes of discontinuities in elastic moduli at $T_c$ for one and two-component OPs, along with the particular degree of freedom that causes the discontinuity (given in parentheses after the expression for the discontinuity). For a one-component OP, only compressional moduli $c_{A_g,i}$ show a discontinuity, caused by fluctuations in the order parameter amplitude $\eta$. For a $B_{2u}+iB_{3u}$ two-component OP, compressional moduli and the $c_{B1g}$ elastic modulus are allowed to show a discontinuity. The discontinuity in the compressional moduli is due to fluctuations in the absolute amplitdue $\eta$ of the OP, as well as fluctuations in the relative amplitude $\theta$ between individual components. The discontinuity in $c_{B1g}$, however, is caused by fluctuations of the relative phase $\gamma$ between different order parameter components.}
	\label{table:elastic moduli jumps for different order parameters}
\end{table}

\subsubsection{Order Parameter Free Energy and Coupling to Strain:\\Two-Component Order Parameter}
Next we discuss discontinuities in the elastic moduli with a two-component OP $\boldsymbol{\eta} = \{ \eta_x, \eta_y \}$.
In the $D_{2h}$ point group, all irreducible representations are one dimensional. A two-component order parameter therefore has to consist of two one-component order parameters, meaning $\eta_x$ and $\eta_y$ can belong to different irreducible representations and are not related by symmetry. The example of $\eta_x$ and $\eta_y$ transforming as the $B_{2u}$ and $B_{3u}$ irreducible representations, respectively, as suggested for the superconducting OP in \ute by authors in \citet{hayes2021multicomponent} and \citet{weiInterplayMagnetismSuperconductivity2022}, will be used in the discussion below. For this particular OP, three independent bilinear combinations can be formed: $\left|\eta_x\right|^2$, $\left|\eta_y\right|^2$, and $\left(\eta_x \eta_y^* + \eta_x^* \eta_y \right)$ transforming as $A_g$, $A_g$, and $B_{1g}$ representations respectively. The Landau free energy reads
\begin{align}
    \label{eq:OP free energy 2 component}
    f_{OP} &= \frac{a_1}{2} \left( \left|\eta_x\right|^2 + \left|\eta_y\right|^2 \right) + \frac{a_2}{2} \left( \left|\eta_x\right|^2 - \left|\eta_y\right|^2 \right)\\
    \notag
    &+ \frac{b_1}{4} \left(\left|\eta_x\right|^4 + \left|\eta_y\right|^4\right) + \frac{b_2}{4} \left(\left|\eta_x\right|^4 - \left|\eta_y\right|^4\right) + \frac{b_3}{2} \left|\eta_x\right|^2 \left|\eta_y\right|^2 + \frac{b_4}{4} \left( \left(\eta_x \eta_y^*\right)^2 + \left(\eta_x^* \eta_y\right)^2 \right),
\end{align}
where $a_{1,2} = a_{1,2}^{(0)} \left(T-T_c\right)$, $a_{1,2}>0$, and $b_i$ are phenomenological constants. Based on these considerations, the free energy coupling the OP to linear powers of strain can be written as
\begin{align}
    \label{eq:coupling free energy 2 component}
    f_{coupling} &= \frac{1}{2} \left( g_1^s \varepsilon_{Ag,1} + g_2^s \varepsilon_{Ag,2} + g_3^s \varepsilon_{Ag,3} \right) \left( \left|\eta_x\right|^2 + \left|\eta_y\right|^2 \right) + \frac{1}{2} \left( g_1^a \varepsilon_{Ag,1} + g_2^a \varepsilon_{Ag,2} + g_3^a \varepsilon_{Ag,3} \right) ]\left( \left|\eta_x\right|^2 - \left|\eta_y\right|^2 \right)\\
    \notag
    &+ \frac{g_4}{2} \varepsilon_{B1g} \left(\eta_x \eta_y^* + \eta_x^* \eta_y \right).
\end{align}
Coupling of $B_{1g}$ strain to the second power of the OP as in the free energy above is only possible for the particular example of a $\{B_{2u},B_{3u}\}$ OP. However, linear coupling of shear strain (i.e. $B_{1g}$, $B_{2g}$, or $B_{3g}$ strain in $D_{2h}$) to a bilinear of the OP is in general only possible for a two-component OP.

In order to calculate the discontinuities of elastic moduli in the presence of a two-component OP, \citet{ghosh_thermodynamic_2021} generalized the expression in \autoref{eq:jump in el mod for one component OP} to
\begin{equation}
    \label{eq:jump in el mod for two component OP}
    \delta c_{mn} = \left. - \boldsymbol{Z}_m^T \boldsymbol{Y}^{-1} \boldsymbol{Z}_n \right|_{\boldsymbol{\eta}\rightarrow\boldsymbol{\eta}_0,\varepsilon_m \rightarrow 0},
\end{equation}
where $\boldsymbol{Z}_m = \frac{\partial^2 f_{coupling}}{\partial \boldsymbol{\eta} \partial \varepsilon_m} $ and $\boldsymbol{Y} = \frac{\partial^2 f_{OP}}{\partial \boldsymbol{\eta}^2}$. Parametrizing the OP as $\boldsymbol{\eta} = \eta \left\{ \cos\theta, e^{i\gamma}\sin\theta \right\}$, the derivative $\partial/\partial \boldsymbol{\eta}$ becomes $\partial/\partial \left\{\eta, \theta, \gamma \right\}$. Assuming a chiral order parameter $\left( \theta_0, \gamma_0 \right) = \left( \pi/4, \pm \pi/2 \right)$, the equilibrium amplitude $\eta_0$, defined by $\left. \partial f_{OP} / \partial \eta \right|_{\eta_0,\theta_0,\gamma_0}=0$, is then given by $\eta_0 = \pm \sqrt{\frac{-2 a_1}{b_1+b_3-b_4}}$. This assumption is motivated by the observation of time-reversal symmetry breaking (TRSB) \cite{hayes2021multicomponent,weiInterplayMagnetismSuperconductivity2022} in the superconducting state of \ute. For this order parameter configuration, one finds
\begin{align}
    \label{eq:Y matrix}
    \boldsymbol{Y} = 
	\begin{pmatrix}
        a_1 + \frac{3\eta_0^2}{2}\left(b_1+b_3-b_4\right) & -2 \eta_0 \left(a_2 + b_2\eta_0^2\right) & 0\\
	    -2 \eta_0 \left(a_2 + b_2\eta_0^2\right)          & \left( b_1-b_3+b_4\right)\eta_0^4        & 0\\
	    0                                                 & 0                                        & \frac{b_4 \eta_0^4}{2}\\
	\end{pmatrix},\\
    \label{eq:Z vectors}
	\boldsymbol{Z}_{Ag,i} = 
	\begin{pmatrix}
        g_i^s \eta_0\\
	    -g_i^a \eta_0^2 \\
	    0 \\
	\end{pmatrix},~  
    \boldsymbol{Z}_{B1g} = 
	\begin{pmatrix}
        0 \\
	    0 \\
	    -\frac{g_4 \eta_0^2}{2} \\
    \end{pmatrix},~  
    \boldsymbol{Z}_{B2(3)g} = 
	\begin{pmatrix}
        0 \\
	    0 \\
	    0 \\
    \end{pmatrix},
\end{align}
where $i = 1,2,3$. From \autoref{eq:Y matrix} and \autoref{eq:Z vectors} in can be seen that for a chiral $\left\{ B_{2u}, B_{3u} \right\}$ order parameter in a $D_{2h}$ point group, all compressional moduli (i.e. the elastic moduli corresponding to $A_g$ strains) show a step discontinuity at $T_c$ due to coupling of the corresponding strain to the absolute amplitude of the OP (the superfluid density), as well as the relative amplitude of the different components (this is in contrast to a multi-component OP where the different components are related by symmetry, for which compressional strains only couple to the absolute amplitude of the OP \cite{ghosh_thermodynamic_2021}). Among all the shear moduli, only $c_{B1g}$ shows a step discontinuity at $T_c$, due to the coupling of $B_{1g}$ shear strain to the relative phase between the different components of the OP.

While the details of the above calculation depend on the exact OP parameter configuration, the main statement is general: a multi-component OP is required for a discontinuity in any shear modulus.
\vfill

\pagebreak
\subsection{Heat Capacity Measurements}
Heat capacity measurements (\autoref{fig:heat capacity}) were performed in a $^3$He cryostat using the quasi-adiabatic method: a fixed power was applied to the calorimeter to raise it approximately 1\% over the bath temperature. The power was then turned off to allow the calorimeter to relax back to the bath temperature. The heat capacity was extracted from these heating and cooling data by fitting them to exponentially-saturating curves. The sample was affixed to the calorimeter with Apiezon N grease. The background heat capacity of the grease and the calorimeter were measured separately and subtracted from the data in \autoref{fig:heat capacity}.

\subsection{Ehrenfest Analysis}
\begin{figure}[h]%
    \includegraphics[width=.5\columnwidth]{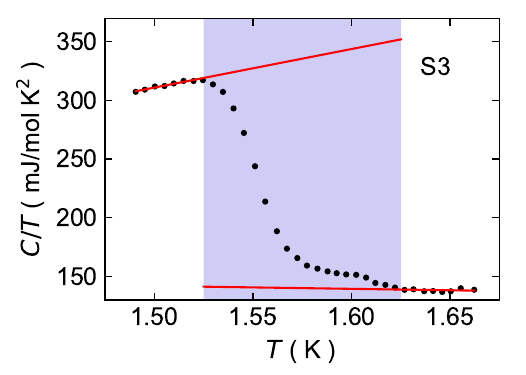}%
    \caption{\textbf{Heat Capacity.} Heat capacity divided by temperature as a function of temperature measured on sample S3. The jump at $T_c$ is $(196\pm18)$~mJ/(mol\,K$^2$), determined according to linear fits below and above the transition (red lines). The uncertainty is estimated from the finite temperature range close to $T_c$ in which the data deviates significantly from these fits. This range is indicated by the blue shaded region.}%
    \label{fig:heat capacity}%
\end{figure}

\begin{figure}%
    \includegraphics[width=1\columnwidth]{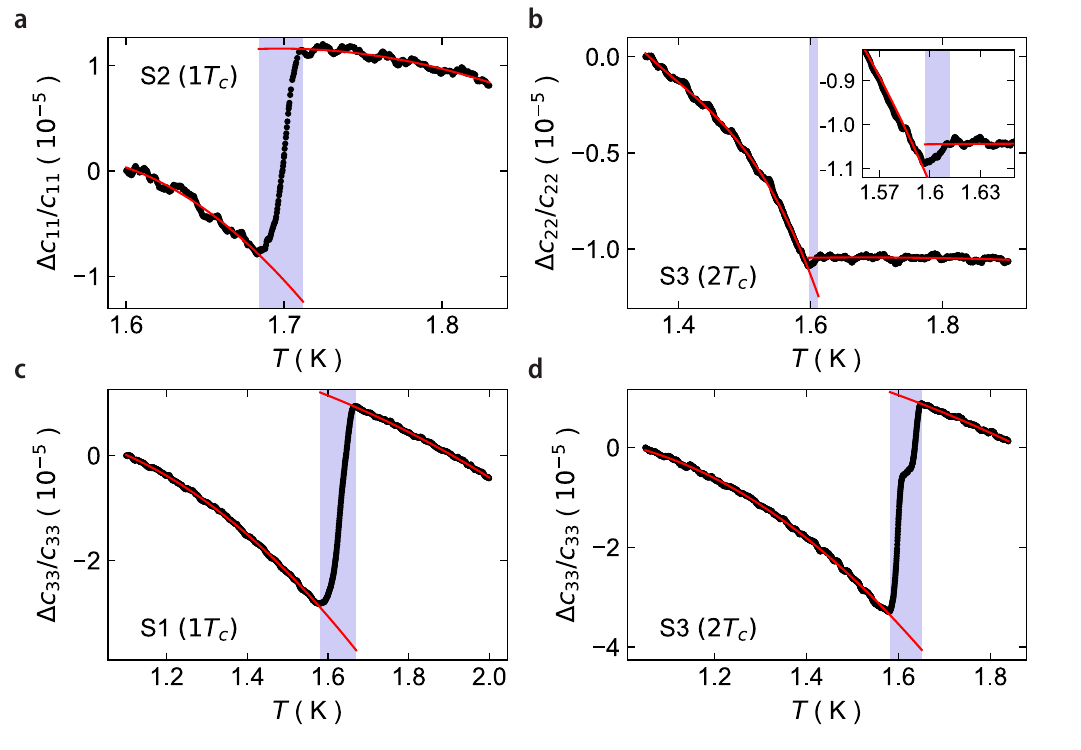}%
    \caption{\textbf{Step Discontinuities in Compressional Moduli.} Relative changes in compressional moduli across $T_c$ (black points). The magnitude of the step discontinuity in compressional moduli is defined as the difference between polynomial fits to the data above and below $T_c$ (red lines). The uncertainty is estimated from the finite temperature range close to $T_c$ in which the data deviates significantly from these fits (indicated in blue). Extracted values are given in \autoref{table:Ehrenfest}. The elastic moduli shown are $c_{11}$ (panel a) and $c_{33}$ (panel c) for samples with one transition and $c_{22}$ (panel b) and $c_{33}$ (panel d) for samples with two transitions. The inset of panel b shows the relative change in $c_{22}$ close to $T_c$.}%
    \label{fig:Ehrenfest}%
\end{figure}

The discontinuity observed in the compressional moduli $\delta c_{ii}$ at $T_c$ is directly related to the jump in the heat capacity divided by temperature, $\Delta C/T$, via Ehrenfest relations. For a single component order parameter they read
\begin{equation}
    \label{eq:Ehrenfest compressional moduli}
    \delta c_{ii} = - \frac{\Delta C}{T} \left( \frac{d T_c}{d \varepsilon_{ii}} \right)^2.
\end{equation}
The derivative of critical temperature with respect to compressional strain $dT_c/d\varepsilon_{ii}$ can therefore be calculated by extracting the discontinuities of the corresponding elastic modulus and the heat capacity at $T_c$.
The heat capacity was measured in sample S3 (see \autoref{fig:heat capacity}) and the size of its discontinuity at $T_c$ is determined to be $(196\pm18)$~mJ/(mol\,K$^2$). The magnitudes of the discontinuities in $\Delta c/c$ for all compressional moduli are extracted according to \autoref{fig:Ehrenfest} and the values are given in \autoref{table:Ehrenfest}. Using these values, as well as the elastic moduli of \ute \cite{theuss2023resonant}, the absolute values of $dT_c/d\varepsilon_{ii}$ ($ii = xx, yy, zz$) are calculated (see \autoref{table:Ehrenfest}).

\begin{table}[b]
	% \centering
	% \begin{tabular}{cccccc}
	% 	Elastic Modulus   & Step in $\frac{\Delta c}{c}$ & $\left|\frac{d T_c}{d\varepsilon_{ii}}\right|$ ( $\frac{\mathrm{K}}{\mathrm{\%\,strain}}$ ) & $\frac{1}{c_{ii}}\left|\frac{d T_c}{d\varepsilon_{ii}}\right|$ ( $\frac{\mathrm{K}}{\mathrm{GPa}}$ ) & $\left|\frac{d T_c}{d\sigma_{ii}}\right|$ ( $\frac{\mathrm{K}}{\mathrm{GPa}}$ ) & $\frac{d T_c}{d\sigma_{ii}}$ ( $\frac{\mathrm{K}}{\mathrm{GPa}}$ ) from \cite{girod2022ThermodynamicElectricalTransport}\\
	% 	\toprule
    %     $c_{11}$ ($1 T_c$) & $-(2.2\pm0.2)\times 10^{-5}$   &  $0.23\pm0.01$   & $0.26\pm0.02$ & $0.51\pm0.02$ & -0.87\\
    %     $c_{22}$ ($2 T_c$) & $-(0.10\pm0.09)\times 10^{-5}$ &  $0.06\pm0.03$   & $0.04\pm0.02$ & $0.09\pm0.02$ & -0.45\\
    %     $c_{33}$ ($1 T_c$) & $-(4.3\pm0.3)\times 10^{-5}$   &  $0.34\pm0.02$   & $0.35\pm0.02$ & $0.60\pm0.03$ & 0.56\\
    %     $c_{33}$ ($2 T_c$) & $-(4.7\pm0.2)\times 10^{-5}$   &  $0.35\pm0.02$   & $0.37\pm0.02$ & $0.62\pm0.03$ & 0.56\\	
	% 	\bottomrule
	% \end{tabular}
    \begin{tabular}{ccccc}
		Elastic Modulus   & Step in $\frac{\Delta c}{c}$ & $\frac{d T_c}{d\varepsilon_{ii}}~\left(\frac{\mathrm{K}}{\mathrm{\%\,strain}}\right)$ & $\frac{d T_c}{d\sigma_{ii}}~ \left(\frac{\mathrm{K}}{\mathrm{GPa}}\right)$ & $\frac{d T_c}{d\sigma_{ii}}~ \left(\frac{\mathrm{K}}{\mathrm{GPa}}\right)$ from \cite{girod2022ThermodynamicElectricalTransport}\\
		\toprule
        $c_{11}$ ($1 T_c$) & $-(2.2\pm0.2)\times 10^{-5}$   &  $-0.23\pm0.02$   & $-0.50\pm0.03$ & -0.87\\
        $c_{22}$ ($2 T_c$) & $-(0.13\pm0.07)\times 10^{-5}$ &  $-0.07\pm0.02$   & $-0.09\pm0.02$ & ---\\
        $c_{33}$ ($1 T_c$) & $-(4.4\pm0.3)\times 10^{-5}$   &  $0.34\pm0.02$   & $0.60\pm0.03$ & 0.56\\
        $c_{33}$ ($2 T_c$) & $-(4.7\pm0.2)\times 10^{-5}$   &  $0.35\pm0.02$   & $0.62\pm0.03$ & 0.56\\	
		\bottomrule
	\end{tabular}
	\caption{\textbf{Ehrenfest Analysis.} Derivatives of the critical temperature with respect to strain $dT_c/d\varepsilon$ are calculated based on the magnitudes of the discontinuities in $\Delta c/c$ extracted according to \autoref{fig:Ehrenfest}. Values of the absolute elastic moduli and respective uncertainties are taken from \citet{theuss2023resonant} and the size of the specific heat jump (or more precisely $\Delta C/T$) is taken to be $196\pm18$~mJ/(mol\,K$^2$) from \autoref{fig:heat capacity}. Knowledge of the signs of $dT_c/d\varepsilon$ is required for the correct calculation of $dT_c/d\sigma$. Since \ref{eq:Ehrenfest compressional moduli} only yields their absolute value, the signs are chosen according to uniaxial stress experiments \cite{girod2022ThermodynamicElectricalTransport}. The last column shows a comparison to values from said uniaxial stress experiments in \citet{girod2022ThermodynamicElectricalTransport}.}
	\label{table:Ehrenfest}
\end{table}

The derivatives of the critical temperature with respect to stress can be calculated from the derivatives with respect to strain via
\begin{equation}
    \label{eq:dTcdsigma}
    \begin{pmatrix}
        dT_c/d\sigma_{xx}\\
	    dT_c/d\sigma_{yy}\\
	    dT_c/d\sigma_{zz}\\
    \end{pmatrix}
    = 
    \begin{pmatrix}
        c11 & c12 & c13\\
	    c12 & c22 & c23\\
	    c13 & c23 & c33\\
	\end{pmatrix}^{-1}
    \begin{pmatrix}
        dT_c/d\varepsilon_{xx}\\
	    dT_c/d\varepsilon_{yy}\\
	    dT_c/d\varepsilon_{zz}\\
    \end{pmatrix}.
\end{equation}
The resulting values are given in \autoref{table:Ehrenfest}, along with values measured in uniaxial stress experiments \cite{girod2022ThermodynamicElectricalTransport}. The elastic tensor used for this calculation is again taken from \citet{theuss2023resonant}. Note that the analysis in \autoref{eq:dTcdsigma} requires knowledge about the signs of $dT_c/d\varepsilon_{ii}$, whereas the Ehrenfest relations in \autoref{eq:Ehrenfest compressional moduli} only yield their absolute values. For a correct analysis from our data, signs according to \citet{girod2022ThermodynamicElectricalTransport} were assumed.
\vfill

\clearpage
\subsection{\ute Fermi Surface and Superconducting Gap}
\subsubsection{Density Functional Theory}

\begin{figure}%
\includegraphics[width=\columnwidth]{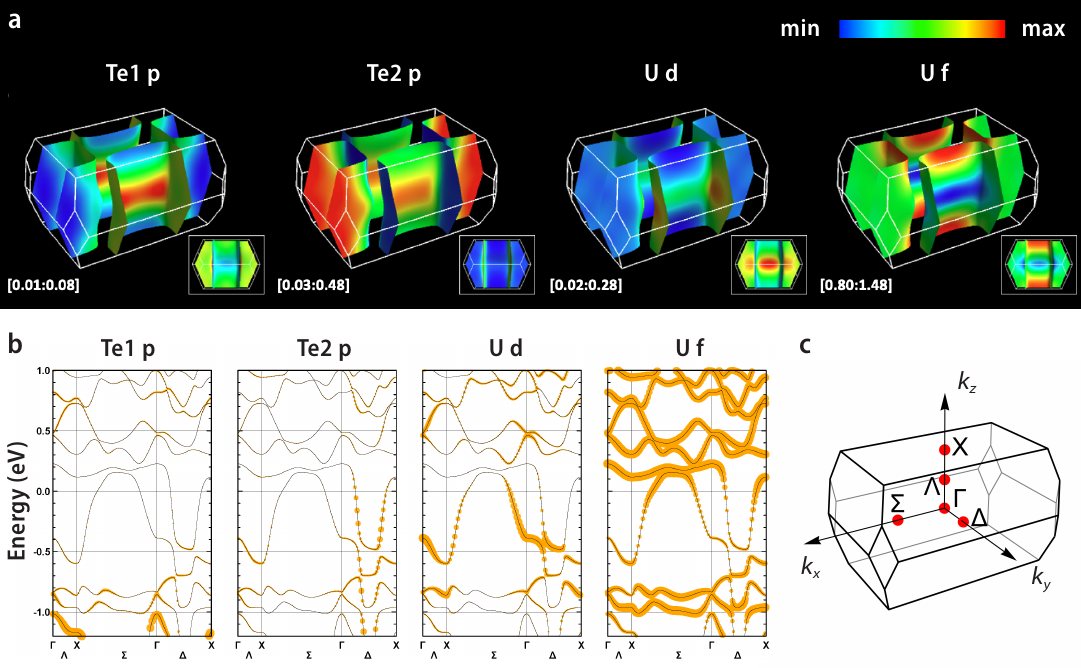}%
\caption{\textbf{DFT Fermi Surface and Band Structure.} (a) Fermi surface of \ute calculated with $U=2$~eV. Fermi surfaces are colored according to their orbital Te1 $p$, Te2 $p$, U $d$, and U $f$ content, from left to right, respectively. Color scales are rescaled between each plot, but respective minimum and maximum values are given to the bottom left of each panel. (b) Band structure calculated with the same parameters as in (a). Orange circles are sized according to the orbital weight of Te1 $p$, Te2 $p$, U $d$, and U $f$ (from left to right) on each individual band. (c) \ute Brillouin zone.}%
\label{fig:DFT}%
\end{figure}

Density-functional theory calculations are used to examine the orbital character of the electronic states in the vicinity of the chemical potential. The self-consistent field calculation is performed in the same way as in \citet{theuss2023resonant}, by additionally considering the Hubbard $U$ for the uranium $5f$ electrons. The full-potential linearized augmented plane wave method \cite{weinertFLAPWApplicationsImplementations2009} calculations employed the generalized gradient approximation \cite{PBE} for the exchange correlation, wave function and potential energy cutffs of 16 and 200 Ry, respectively, and muffin-tin sphere radii of $\SI{1.35}{\angstrom}$. Spin-orbit coupling was fully taken into account in the assumed nonmagnetic state. We set $U=2$~eV to obtain a quasi 2D Fermi surface \cite{ishizuka2019InsulatorMetalTransitionTopological,shishidou2021TopologicalBandSuperconductivity}, which qualitatively accounts for the recent experiments. Along the high-symmetry lines in the Brillouin zone ($\Lambda$, $\Sigma$, and $\Delta$ lines, see \autoref{fig:DFT}c) and on a dense 50$\times$50$\times$50 $k$-point mesh, the (Kramers degenerate) band energy and wave functions are generated, and the orbital components of each doublet are calculated within the atom-centered spheres of radius $\SI{1.35}{\angstrom}$. In \autoref{fig:DFT}, the orbital components are shown on the Fermi surface (panel a---the visualization of the Fermi surface is done with \texttt{FermiSurfer} \cite{KAWAMURA2019197}) and along the band dispersion (panel b).

% DFT calculations of \ute were carried out by the full-potential linearized augmented plane wave method \cite{weinertFLAPWApplicationsImplementations2009}. The experimentally determined lattice parameters \cite{ikedaSingleCrystalGrowth2006} are used: space group $Immm$; $a=\SI{4.1611}{\angstrom}$, $b=\SI{6.1222}{\angstrom}$, $c=\SI{3.955}{\angstrom}$ with fractional coordinates (0, 0, 0.13544), (0.5, 0, 0.29750), (0, 0.2509, 0.5) for the U $4i$ site and Te $4j$ and $4h$ sites, respectively. The muffin-tin sphere radii are set to $\SI{1.35}{\angstrom}$, and the wave function and potential cutoffs are 16 and 100 Ry, respectively. The Brillouin zone is sampled with a $15\times 15\times 15$ $k$-point mesh during the self-consistent field cycle. To account for the Coulomb correlation in U $5f$ states, a rotationally invariant version of the local density approximation plus Hubbard U (LDA+U) method \cite{yareskoLocalized5fElectrons2003} is employed, with the full $14\times 14$ $5f$ occupation matrix and the Slater integrals other than the monopole term $F_0 = U$ set to zero \cite{ishizuka2019InsulatorMetalTransitionTopological}. Fermi surface and band structure plots in \autoref{fig:DFT} are performed with $U=2$~eV.

\subsubsection{Tight Binding Model}

\autoref{fig:DFT} motivates a tight binding model constructed from two quasi-one-dimensional chain Fermi surfaces: one chain from the Te(2) 5$p$ orbitals, and one from the U 6$d$ orbitals. This model faithfully captures the shape of the Fermi surface measured by quantum oscillations (see \cite{eaton2023Quasi2DFermiSurface}). This Fermi surface is quite similar to that calculated for ThTe$_2$, which has no $f$ electrons---while the $f$ electrons in \ute hybridize strongly with both bands, the predominant effect is to enhance the cyclotron masses and shift the chemical potential, without strongly modifying the Fermi surface shape.

There are two uranium atoms that form a dimer in the center of the conventional unit cell shown in Figure 5a of the main text. The dominant tight binding parameters will be the chemical potential $\mu_{\rm U}$, the intra-dimer overlap $\Delta_{\rm U}$, the hopping $t_{\rm U}$ along the uranium chain in the $a$ direction, the hopping $t'_{\rm U}$ to other uranium in the dimer along the chain direction, the hoppings $t_{ch,{\rm U}}$ and $t'_{ch,{\rm U}}$ between chains in the $a-b$ plane, and the hopping $t_{z,{\rm U}}$ between chains along the $c$ axis. The two bands from the two uranium sites then come from diagonalizing the following matrix:
\begin{equation}
E_{\rm U} = {\tiny
\begin{bmatrix}
 \mu_{\rm U} - 2t_{\rm U} \cos k_x a - 2t_{ch,{\rm U}}\cos k_y b  & -\Delta_{\rm U} - 2t'_{\rm U} \cos k_x a - 2t'_{ch,{\rm U}}\cos k_y b -4t_{z,{\rm U}} e^{-i k_z c/2}\cos k_x \frac{a}{2}\cos k_y \frac{b}{2} \\
 -\Delta_{\rm U} - 2t'_{\rm U} \cos k_x a - 2t'_{ch,{\rm U}}\cos k_y b -4t_{z,{\rm U}} e^{i k_z c/2}\cos k_x \frac{a}{2}\cos k_y \frac{b}{2} & \mu_{\rm U} - 2t_{\rm U} \cos k_x a - 2t_{ch,{\rm U}}\cos k_y b 
\end{bmatrix}
}.
\label{eq:EU}
\end{equation}

There are in principle 4 Te(2) sites per conventional unit cell, but by including only nearest-neighbor hopping in the $a-b$ plane, the problem is again reduced to diagonalizing a $2\times 2$ matrix. The dominant tight binding parameters are then the chemical potential $\mu_{\rm Te}$, the intra-unit-cell overlap $\Delta_{\rm Te}$ between the two Te(2) atoms along the chain direction, the hopping $t_{\rm Te}$ along the Te(2) chain in the $b$ direction, the hopping $t_{ch,{\rm Te}}$ between chains in the $a$ direction, and the hopping $t_{z,{\rm Te}}$ between chains along the $c$ axis. The tight binding matrix is:
\begin{equation}
E_{\rm Te} = {\tiny
\begin{bmatrix}
 \mu_{\rm Te} -2t_{ch,{\rm Te}} \cos k_x a & -\Delta_{\rm Te}-t_{\rm te} e^{-i k_y b} - 2 t_{z{\rm,Te}}\cos k_z \frac{c}{2} \cos k_x \frac{a}{2}\cos k_y \frac{b}{2} \\
 -\Delta_{\rm Te}-t_{\rm te} e^{i k_y b} - 2 t_{z{\rm,Te}}\cos k_z \frac{c}{2} \cos k_x \frac{a}{2}\cos k_y \frac{b}{2} & \mu_{\rm Te} -2t_{ch,{\rm Te}} \cos k_x a
\end{bmatrix}
}.
\label{eq:ETe}
\end{equation}

\begin{figure}%
\includegraphics[width=1\columnwidth]{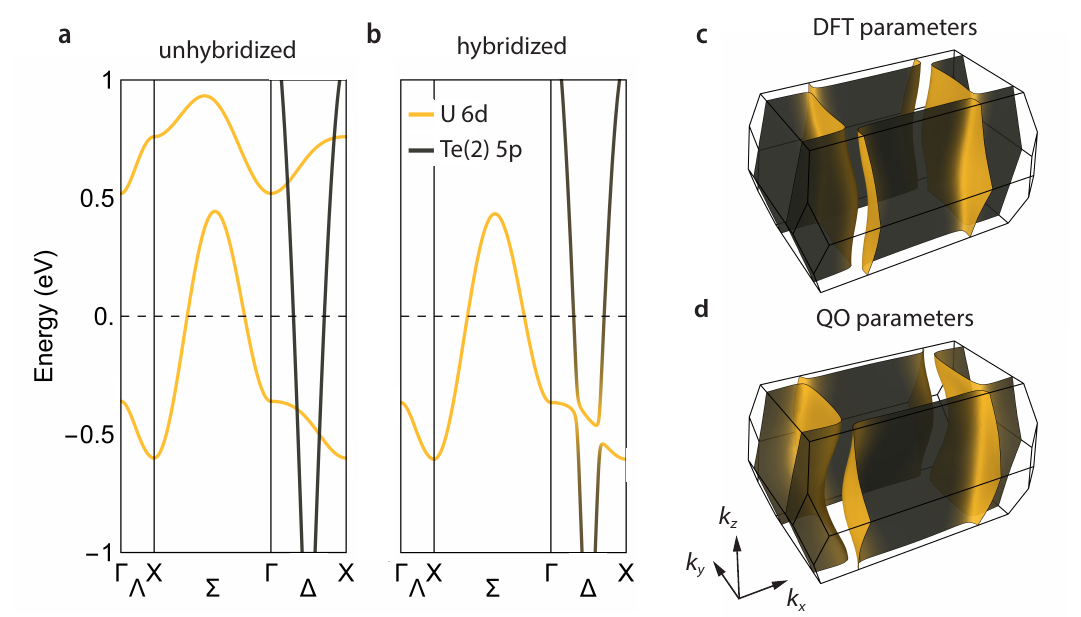}%
\caption{\textbf{Tight Binding Model.} (a-c) Tight binding model with parameters to match DFT results. (a) Unhybridized bands formed by uranium $6d$ (yellow) and tellurium (2) $5p$ (gray) electrons. Bands crossing the Fermi level are hybridized in (b) and the resulting Fermi surface is shown in (c). Colors represent a projection on original U/Te(2) bands. (d) Fermi surface from tight binding model with parameters to match quantum oscillation results \cite{eaton2023Quasi2DFermiSurface}.}%
\label{fig:tightDFT}%
\end{figure}

The resultant bands are plotted in \autoref{fig:tightDFT}a. The tight binding parameters were chosen to roughly match the DFT result shown in \autoref{fig:DFT} and are given in \autoref{tab:binding}. The two bands crossing the Fermi energy can be hybridized to form the electron and hole pocket. We use an isotropic in momentum hybridization $\delta$ and chose its value to roughly match the DFT result. The resultant two bands that cross the Fermi energy are shown in \autoref{fig:tightDFT}b, and the 3D Fermi surface is shown in \autoref{fig:tightDFT}c. The predominant difference between the FS calculated with $U = 2$~eV and the FS reported by \citet{eaton2023Quasi2DFermiSurface} is that the latter was chosen with the opposite-sign dispersion along the $c$-axis. Tight binding parameters chosen to roughly match the FS reported in \citet{eaton2023Quasi2DFermiSurface} are also given in \autoref{tab:binding}, with the resultant FS shown in \autoref{fig:tightDFT}d.

\begin{table}%
\begin{tabular}{|l|c|c|c|c|c|c|c|c|c|c|c|c|}
\hline
&$\Delta_{\rm U}$& $t_{\rm U}$ & $t'_{\rm U}$ & $t_{ch,{\rm U}}$ & $t'_{ch,{\rm U}}$ & $t_{z,{\rm U}}$ & $\mu_{\rm Te}$ & $\Delta_{\rm Te}$  & $t_{\rm Te}$& $t_{ch,{\rm Te}}$ & $t_{z,{\rm Te}}$ & $\delta$ \\
\hline
DFT &0.40 & 0.15 & 0.08 & 0.01 & 0.00 & -0.03 & -1.80 & -1.50 & -1.50 & 0.00 & -0.05 & 0.09\\
QO &0.05 & 0.10 & 0.08 & 0.01 & 0.00 & 0.04 & -1.80 & -1.50 & -1.50 & -0.03 & -0.5 & 0.10\\
\hline
\end{tabular}
\caption{All parameters given in eV.}
\label{tab:binding}
\end{table}

\subsubsection{Superconducting Gap}
When considering the symmetries of superconducting gaps, it is necessary to distinguish the cases of weak and strong spin-orbit coupling: \ute likely falls in the latter category. However, since the orthorhombic point group of \ute ($D_{2h}$) is inversion symmetric, one can still label irreducible representations as even or odd. This classification is used to distinguish spin singlet (even) or triplet (odd) superconductors. Since \ute is most likely a spin-triplet superconductor, the possible irreducible representations of the order parameter are $A_u$, $B_{1u}$, $B_{2u}$, and $B_{3u}$. In the strong spin-orbit limit, they correspond to the following $\vec{d}$-vectors \cite{annett1990SymmetryOrderParameter}
\begin{align}
    \label{eq:Au d vector}
    \vec{d}_{A_u}    &= \left\{ \alpha k_x, \beta k_y, \gamma k_z \right\},\\
    \label{eq:B1u d vector}
    \vec{d}_{B_{1u}} &= \left\{ \alpha k_y, \beta k_x, \gamma k_x k_y k_z \right\},\\
    \label{eq:B2u d vector}
    \vec{d}_{B_{2u}} &= \left\{ \alpha k_z, \beta k_x k_y k_z, \gamma k_x \right\},\\
    \label{eq:B2u d vector}
    \vec{d}_{B_{3u}} &= \left\{ \alpha k_x k_y k_z, \beta k_z, \gamma k_y \right\},
\end{align}
where $\alpha$, $\beta$, and $\gamma$ are real constants and the momentum dependence of the superconducting gap is given by
\begin{equation}
    \label{eq:SC gap function}
    \Delta \left(\vec{k}\right) = \sqrt{\vec{d}\cdot \vec{d}^\star  \pm \left| \vec{d} \times \vec{d}^\star\right| }.
\end{equation}
Here, $\vec{d}^\star$ is the complex conjugate of $\vec{d}$. The $A_u$ order parameter is fully gapped, whereas the $B_{1u}$, $B_{2u}$, and $B_{3u}$ order parameters have point nodes along the $k_z$, $k_y$, and $k_x$ directions respectively. A $B_{1u}$ gap is thus also fully gapped on the Fermi surface of \ute found by quantum oscillations \cite{aoki2022FirstObservationHaas, eaton2023Quasi2DFermiSurface} and only exhibits point nodes on a putative Fermi pocket enclosing the $\Gamma$-point \cite{miaoLowEnergyBand2020}.

The gap structures shown in the main text are calculated at $k_z = 0$ with $\alpha=\beta=\gamma$. A slight anisotropy in these parameters can change the exact shape of the momentum dependence of the different gap symmetries, but will not change their nodal structure.

\bibliography{literature}

\bibliography{literature}

\end{document}